\documentclass[aps,pra,10pt,twocolumn,showpacs]{revtex4-1}

\usepackage{amssymb,amsfonts,amsmath,bm,bbm}
\usepackage{graphicx}

\newcommand{\abs}[1]{\left| #1 \right|}
\newcommand{\ii}{\mathrm{i}}
\newcommand{\ee}[1]{\mathrm{e}^{#1}}
\newcommand{\dd}{\mathrm{d}}
\renewcommand{\vec}[1]{\bm{#1}}
\newcommand{\vectornorm}[1]{\left|\left| #1 \right|\right|}

\newcommand{\erw}[1]{\left< #1 \right>}
\newcommand{\expect}[1]{\left< #1 \right>}
\newcommand{\braket}[2]{\left< #1 \vphantom{#2} \right|
 \left. #2 \vphantom{#1} \right>}
\newcommand{\matrixel}[3]{\left< #1 \vphantom{#2#3} \right|
 #2 \left| #3 \vphantom{#1#2} \right>}

\newcommand{\del}[2]{\frac{\partial #1}{\partial #2}}

\DeclareMathOperator{\Real}{Re}
\DeclareMathOperator{\Imag}{Im}
\DeclareMathOperator{\Tr}{Tr}

\makeatletter
\renewcommand*\env@matrix[1][*\c@MaxMatrixCols c]{%
  \hskip -\arraycolsep
  \let\@ifnextchar\new@ifnextchar
  \array{#1}}
\makeatother

\begin{document}

\title{Variational calculations for anisotropic solitons in dipolar 
Bose-Einstein condensates}
\author{R\"udiger Eichler}
\author{J\"org Main}
\author{G\"unter Wunner}
\affiliation{Institut f\"ur Theoretische Physik 1, Universit\"at Stuttgart, 
 70550 Stuttgart, Germany}
\date{\today}

\begin{abstract}
  We present variational calculations using a Gaussian
  trial function to calculate the ground state of the Gross-Pitaevskii
  equation and to describe the dynamics of the quasi-two-dimensional
  solitons in dipolar Bose-Einstein condensates. Furthermore we extend
  the ansatz to a linear superposition of Gaussians improving the
  results for the ground state to exact agreement with numerical grid
  calculations using imaginary time and split-operator method. We are
  able to give boundaries for the scattering length at which stable
  solitons may be observed in an experiment. By dynamical calculations
  with coupled Gaussians we are able to describe the rather complex
  behavior of the thermally excited solitons. The discovery of
  dynamically stabilized solitons indicates the existence of such BECs
  at experimentally accessible temperatures.
\end{abstract}
\maketitle

\section{Introduction}
Since the prediction and the experimental realization of Bose-Einstein
condensates (BECs) the field of cold atomic gases has been subject of
multiple theoretical and experimental investigations. BECs with
long-ranged interaction which can experimentally be realized by the
condensation of atoms with a magnetic dipole moment such as
${}^{52}\mathrm{Cr}$ are of special interest
\cite{Griesmaier05a,Beaufils08,Lahaye09}. To stabilize such
condensates, in general optical traps are applied in all three spatial
directions. However, it has been shown by Tikhonenkov et al.\
\cite{Tikhonenkov08a} that stable quasi-2d solitons are possible where
a trap is applied only in one direction perpendicular to the axis of
the aligned atomic dipoles.

Solitons are a nonlinear effect which arises from the dispersion and
the nonlinearity cancelling out each other. The solitons suffer from
two kinds of instabilities: First, strong attractive particle interactions
can cause the collapse of the condensate and second, when the interactions
are only weakly attractive or even repulsive the BEC can dissolve in 
the directions where the external trap is open.

A prerequisite for the experimental realization of the solitons is a
detailed theoretical investigation of the parameter ranges where the
condensate is stable.
The relevant parameters are the trap frequency, the scattering length 
of the contact interaction, and the excitation energy which allows for
an estimation of the temperature range where the solitons can exist.
The detailed analysis of the stationary states and the dynamics of
solitons based on extended variational calculations is the objective 
of this article.

In the mean-field approximation the BEC with particle number $N$ is 
described by the extended Gross-Pitaevskii equation (GPE).
Introducing ``natural'' units for mass $m_\mathrm{d}=2m$, action $\hbar$, 
length $a_\mathrm{d} =(m\mu_0\mu^2)/(2\pi\hbar^2)$, 
energy $E_\dd=\hbar^2/(2ma_\dd^2)$,
frequency $\gamma_\dd=\hbar/(ma_\dd^2)$, abbreviating
$a=a_\mathrm{sc}/a_\mathrm{d}$ and making use of the scaling properties 
of the GPE 
$(\tilde{\vec{r}},\tilde{\gamma},\tilde{t},\tilde{\psi},\tilde{E}) =
(N^{-1}\vec{r},N^2\gamma , N^{-2}t, N^{3/2}\psi, N E )$ leads to the
scaled extended time-dependent GPE in ``natural'' units
\begin{align}
    \ii \frac{\dd}{\dd t} \psi\left( \vec{r} \right)
    & = \Biggl[ -\Delta +  \gamma_y^2 y^2 
    +8\pi a \abs{\psi\left( \vec{r}\right)}^2 \nonumber\\
    & + \int \dd^3 r'\,
    \frac{1-3\cos^2\theta}{\abs{\vec{r}-\vec{r}'}^3 }\abs{\psi(\vec{r}')}^2
    \Biggr]\psi(\vec{r}) \, ,
\label{eq:GPE}
\end{align}
where the tilde is omitted. 
The dipole moments of the atoms are aligned in $z$ direction by an external
magnetic field, and $\theta$ is the angle between $\vec{r}-\vec{r}'$ and 
the magnetic field axis.
In the trap geometry assumed here only a trap in the $y$-direction 
perpendicular to the magnetic field is present, i.e.\ $\gamma_x=\gamma_z=0$.

In Ref.\ \cite{Tikhonenkov08a} the GPE \eqref{eq:GPE} has been solved
approximately using a variational approach with a Gaussian type
orbital and numerically exact by simulations on a grid.  The grid
calculations are numerically quite expensive.  For a detailed analysis
of the parameter space of the quasi-2d solitons we therefore introduce
and employ an extended variational method based on coupled Gaussian
functions, which has already turned out, for dipolar BECs with an
axisymmetric three-dimensional trap, to be a full-fledged alternative
to grid simulations \cite{Rau10,Rau10a,Rau10b}.  In this article the
stable ground state of the condensate is computed by imaginary time
evolution of an initial wave function, and it is shown that typically
three to six coupled Gaussians are sufficient to obtain fully
converged results.  For a given trap frequency $\gamma_y$ stable
solitons exist in a finite range of the scattering length $a$, outside
that range the condensate collapses or dissolves.

The dynamics of energetically excited solitons is studied by solving
the equations of motions for the variational parameters in real time.
The mean-field energy of the ground state is typically only slightly
below the energy threshold where the soliton can dissolve. However,
the investigation of the dynamics reveals the existence of dynamically
stabilized solitons at energies far above that threshold, indicating
the possible experimental realization of solitons at temperature
$T\approx 5\,\mathrm{\mu K}$ or even higher in the limit of the GPE.
The transition temperature of a chromium BEC is experimentally given
by $T_\mathrm{c}\approx 700\,\mathrm{nK}$ \cite{Griesmaier05a}.
Therefore the soliton can be dynamically stabilized in all such BECs.

The paper is organized as follows:
In Sec.~\ref{chap:singlegauss} we investigate the solitons with the 
ansatz of a single Gaussian wave packet.
The appealing simplicity of this model is that both the stationary states 
and the dynamics of the solitons can be obtained from the Hamiltonian 
of a pseudo particle moving in a three-dimensional potential.
In Sec.~\ref{chap:coupledgauss} the variational ansatz will be extended 
to a linear superposition of Gaussian wave packets (GWPs), and the TDVP 
will be applied to GWPs.
The imaginary time evolution method will be used in 
Sec.~\ref{sec:results_coupled} for the evaluation of the ground state, 
and the results will be compared to the calculations with one Gaussian 
as well as with numerical grid calculations with the split-operator method.
The analysis of the real time dynamics allows us to estimate the stability
of excited solitons at finite temperatures.
Concluding remarks are given in Sec.~\ref{sec:conclusion}.

\section{Variational approach with a single Gaussian}
\label{chap:singlegauss}
Although the variational ansatz with a single Gaussian function cannot
provide quantitatively correct results the simple model already allows us
to gain deep insight into the physics of the quasi-2d solitons.

The ansatz with a single Gaussian as a trial function in the TDVP reads
\begin{align}
 \psi (\vec r) = \ee{\ii \left( A_x x^2 + A_y y^2 + A_z z^2 + \gamma \right)} \,,
\label{eq:1gauss_ansatz}
\end{align}
with the complex variational parameters
\begin{subequations}
\begin{align}
 A_\sigma &= A^r_\sigma + \ii A^i_\sigma \; , \quad \sigma=x,y,z \\
 \gamma &= \gamma^r +\ii \gamma^i \, .
\end{align}
\label{eq:var_param}
\end{subequations}
The parameters $A^i_\sigma$ determine the real half-widths 
$L_\sigma =1/\sqrt{2A^i_\sigma}$ of the Gaussian function, $\gamma^i$ 
describes the amplitude $\hat{\mathcal{A}} = \exp(-\gamma^i)$ used 
for normalization of the wave function, and $\gamma^r$ is a global phase.
The variational ansatz \eqref{eq:1gauss_ansatz} of course strongly
simplifies the problem.
Nevertheless the investigation of this model yields physical insight and the 
results for the ground state are, as will be shown, qualitatively correct.

The mean-field energy and the chemical potential are
\begin{align}
\label{eq:energy_functional}
  E_{\mathrm{mf}} &= \left< -\Delta \right> + \left< V_\mathrm{t} \right> +
  \frac{1}{2} \left(\left< V_\mathrm{c} \right> + \left< V_\mathrm{d}
    \right> \right) \, ,\\
  \varepsilon &= \left< -\Delta \right> + \left< V_\mathrm{t} \right> +
  \left< V_\mathrm{c} \right> + \left< V_\mathrm{d} \right> \, ,
\label{eq:chem_pot}
\end{align}
respectively.
The expectation values with the wave function $\psi$ given 
in Eq.~\eqref{eq:1gauss_ansatz} read
\begin{align}
 \expect{-\Delta} = A_x^i + A_y^i + A_z^i +
\frac{\left( A^r_x \right)^2}{A^i_x} + \frac{\left( A^r_y
\right)^2}{A^i_y} +\frac{\left( A^r_z \right)^2}{A^i_z}
\end{align}
for the kinetic term,
\begin{align}
 \expect{V_\mathrm{t}} = \frac{1}{4} \left(\frac{\gamma_x^2}{A^i_x}
 +\frac{\gamma_y^2}{A^i_y}+\frac{\gamma_z^2}{A^i_z}\right)
\end{align}
for the trapping potential,
\begin{align}
\expect{V_\mathrm{c}} = 8 a \sqrt{\frac{A^i_x A^i_y A^i_z}{\pi}}
\end{align}
for the scattering potential, and
\begin{align}
\expect{V_\mathrm{d}} = \sqrt{\frac{A^i_xA^i_yA^i_z}{\pi^3}}\left(
\frac{4\pi}{3} \left(\kappa_x\kappa_y\,R_D\left(
\kappa_x^2,\kappa_y^2,1 \right) -1\right)\right)
\end{align}
for the dipolar potential with $\kappa_x=\sqrt{A^i_z/A^i_x}$ and
$\kappa_y=\sqrt{A^i_z/A^i_y}$.
The term
\begin{align}
  R_D( x,y,z ) = \frac{3}{2}\int\limits_0^\infty 
  \frac{\dd t}{\sqrt{(x+t)(y+t)(z+t)^3}}
\label{eq:Vd_DEF:R_D}
\end{align}
denotes an elliptic integral of second kind in ``Carlson's form''
which can be evaluated with a fast and stable approximation algorithm
\cite{Carlson94a,Carlson95a} more efficiently than the integral
representation given in \cite{Tikhonenkov08a}.

\subsection{Hamiltonian form of the equations of motion}
\label{sec:eom}
The ground state of the GPE can in principle be obtained by minimizing 
the mean-field energy in Eq.~\eqref{eq:energy_functional}.
Alternatively, the TDVP can be applied to derive equations of motion 
for the variational parameters in Eq.~\eqref{eq:var_param}.
The equations of motion can be used to investigate the dynamics of
the soliton, and the stationary states of the system can be found
by searching for the fixed points of these equations.
The stable fixed point with the lowest mean-field energy denotes the 
ground state.

For the ansatz of a single Gaussian the system can be transformed to the 
descriptive form of a Hamiltonian system by the coordinate transformation
\begin{align}
 A^r_\sigma &= \frac{p_\sigma}{4q_\sigma} \; , \quad
 A^i_\sigma = \frac{1}{8q_\sigma^2} \; , \quad \sigma=x,y,z \, .
\label{eq:canonical}
\end{align}
The Hamiltonian in the $(\vec q,\vec p)$ coordinates
\begin{align}
H&= \frac{p_x^2+p_y^2+p_z^2}{2} +
\frac{1}{8q_x^2} + \frac{1}{8q_y^2}  +\frac{1}{8q_z^2 } +
2\gamma_y^2 q_y^2 \label{eq:hamilton_hamiltonian}\\
& +\frac{\sqrt{\frac{2}{\pi}}a}{8q_xq_yq_z} +
\frac{1}{24 \sqrt{2\pi} q_z}\left(\frac{1}{q_z^2} R_D\left(
\frac{q_x^2}{q_z^2},\frac{q_y^2}{q_z^2},1 \right)-\frac{1}{q_xq_y} \right)
\nonumber
\end{align}
has the conventional form $H=T+V_\mathrm{h}$.
It can be shown that Hamilton's equations 
\begin{align}
 \dot{\vec q} = \frac{\partial H}{\partial{\vec p}} = {\vec p} \; , \quad
 \dot{\vec p} = -\frac{\partial H}{\partial{\vec q}} 
  = -\frac{\partial V_\mathrm{h}}{\partial{\vec q}}
\label{eq:ham}
\end{align}
lead to the same equations of motion as the TDVP applied to the 
variational parameters in Eq.~\eqref{eq:var_param}.

\begin{figure}
\includegraphics[width=0.92\columnwidth]{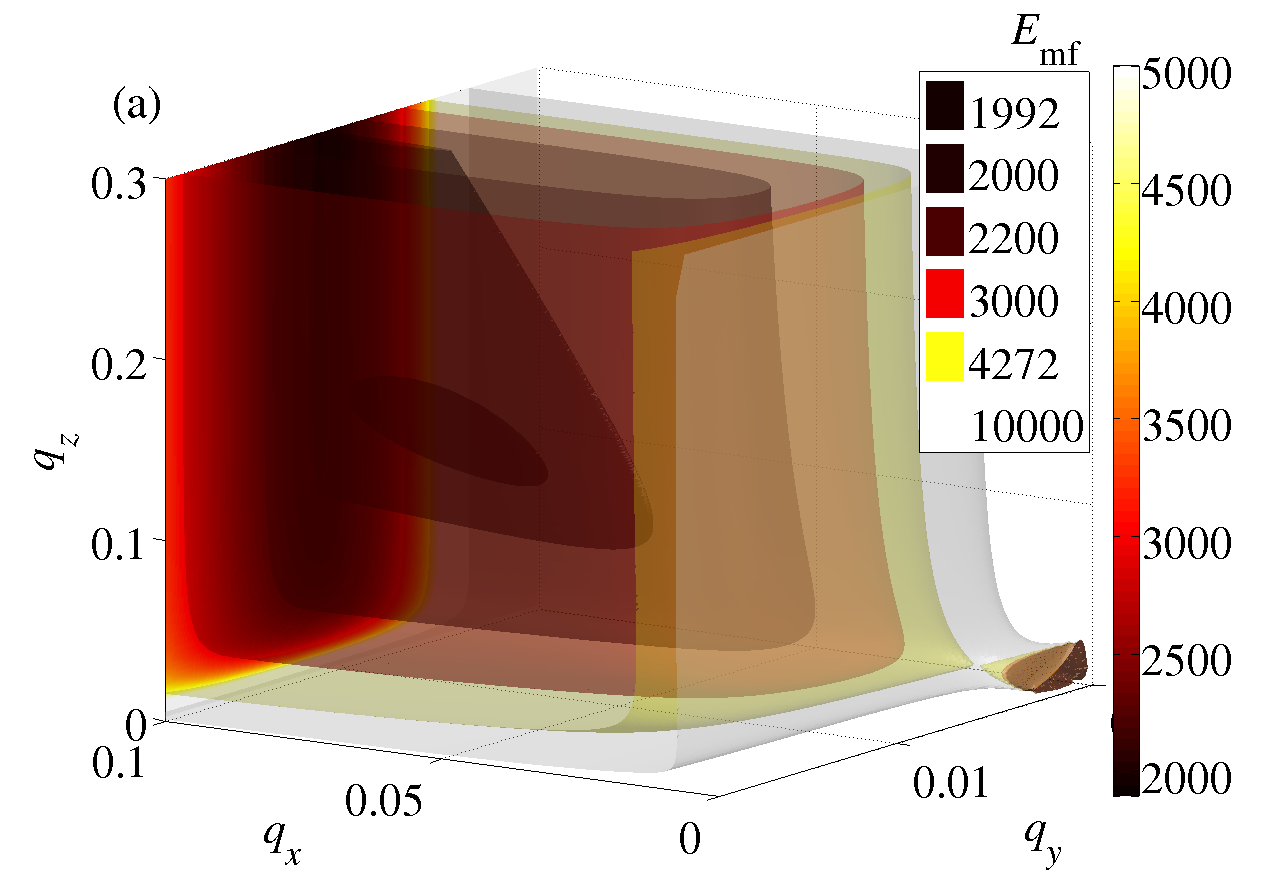}
\includegraphics[width=0.92\columnwidth]{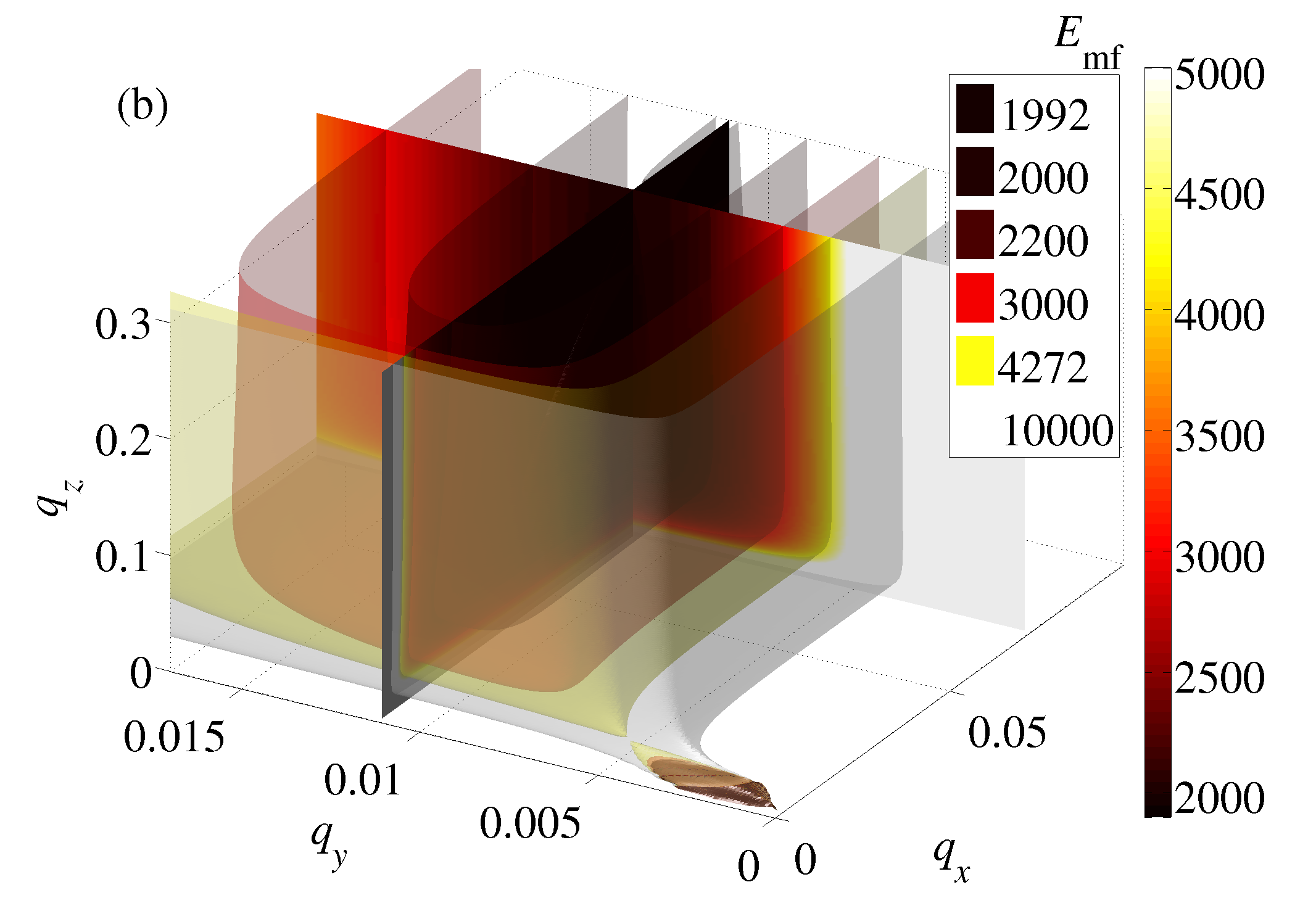}
\includegraphics[width=0.92\columnwidth]{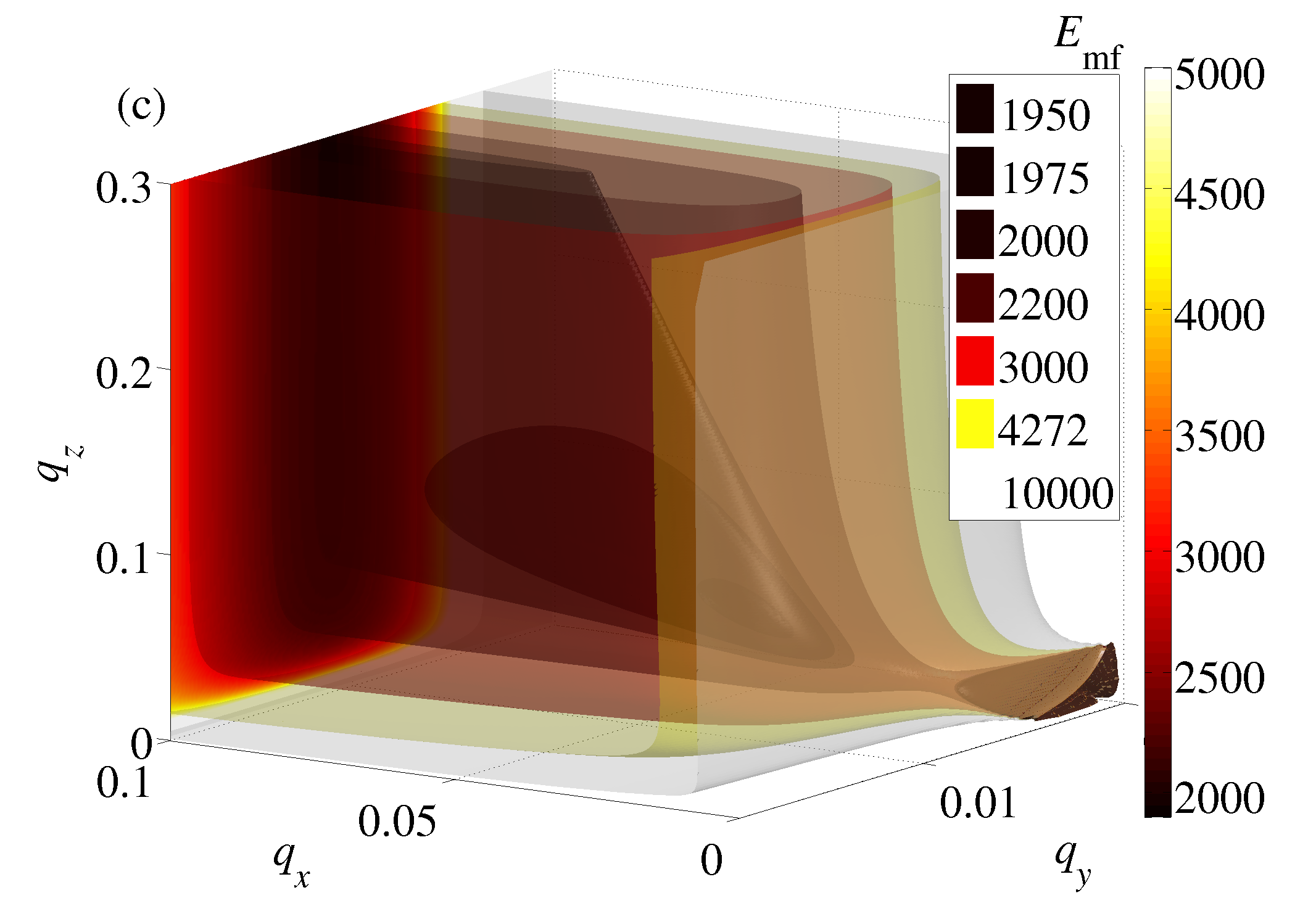}
\caption{(Color online)
 Three-dimensional potential $V_\mathrm{h}(\vec q)$ for $\gamma_y=2000$
 visualized by isosurfaces.  (a) Scattering length $a=0.1$.
 The ellipsoidal form of the isosurfaces marks the stable minimum and the
 hyperbolic form of the isosurfaces close to zero marks the saddle point.
 In (b) the potential is rotated to show that the saddle point lies at
 smaller $q_y$-values than the minimum.
 (c) Same as (a) but at scattering length $a=0.08$ lowered towards the
 bifurcation point. The minimum and the saddle point approach each other.}
\label{fig:potential}
\end{figure}

The potential $V_\mathrm{h}$ is visualized in Fig.~\ref{fig:potential}.
The isopotential surfaces close to the stable fixed point have
ellipsoidal form. At the trap energy $E_\mathrm{mf}=\gamma_y$ the
potential gets open. The unstable fixed point at smaller values of
$q_y$ than the local minimum is given by the saddle point where the
surrounding isosurfaces have hyperbolic form. In
Fig.~\ref{fig:potential}(c) the potential at a smaller scattering
length close to the bifurcation point is shown. The local minimum and
the saddle point approach one another and coincide at the critical 
scattering length $a_\mathrm{crit}$.

\subsection{Stationary states and linear stability}
\label{sec:stat_1g}
\begin{figure}
\includegraphics[width=0.95\columnwidth]{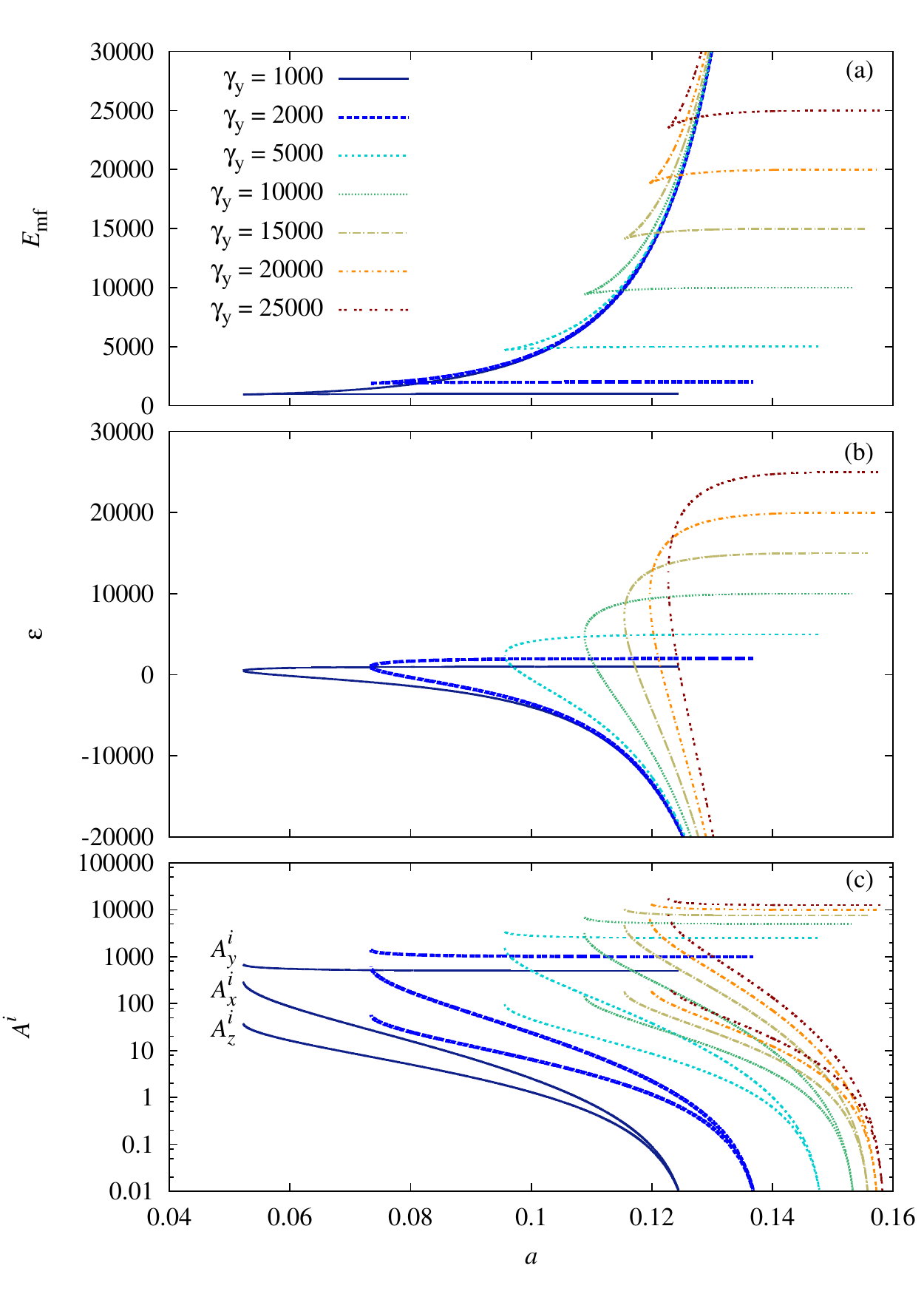}
\caption{(Color online)
 (a) Mean-field energy as function of the scattering
 length for different values of trap frequencies, (b) chemical
 potential and (c) width parameters of the trial function for the
 stable ground state on a logarithmic scale.}
\label{fig:1gauss}
\end{figure}

The fixed points of the system can now be easily calculated by a
nonlinear root search for $\dot{\vec q}=0$ and $\dot{\vec p}=0$
in Hamilton's equations \eqref{eq:ham}.
In Fig.~\ref{fig:1gauss} the mean-field energy and the chemical potential 
of the stable ground state and of an excited unstable state are shown as 
a function of the scattering length. 
The two branches emerge in a tangent bifurcation at the critical 
scattering length $a_\mathrm{crit}$. In
Fig.~\ref{fig:1gauss}(c) the width parameters corresponding to the
ground state show, that with increasing scattering length the soliton
gets very broad and finally dissolves.

\begin{figure}
\includegraphics[width=1\columnwidth]{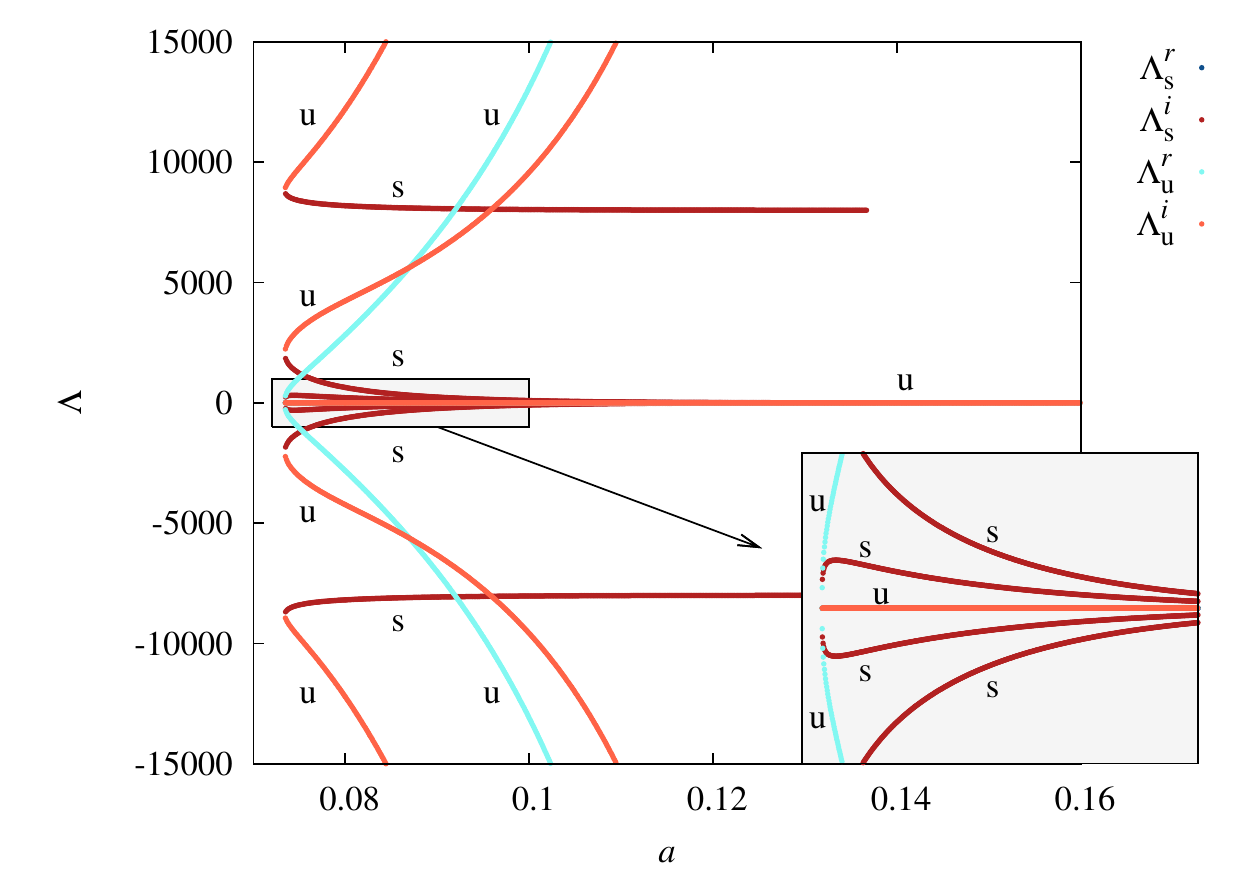}
\caption{(Color online)
 Real and imaginary part of the eigenvalues of the Jacobi matrix $J$
 as a function of the scattering length for the trap frequency
 $\gamma_y=2000$.  The stable branch shows pure imaginary eigenvalues,
 the unstable one has non-vanishing real parts.  The eigenvalues of the
 stable $\Lambda_s$ and unstable branch $\Lambda_u$ coincide at the
 bifurcation point.  The inset shows a magnification of the rectangle.}
\label{fig:stability}
\end{figure}

To analyze the stability of the solution of the GPE the eigenvalues
$\Lambda$ of the six-dimensional Jacobi matrix
\begin{align}
 J=\left. \frac{\partial\left( \dot A^r_\sigma ,\dot
 A^i_\sigma \right)}{\partial \left( A^r_\sigma ,A^i_\sigma
 \right)}\right|_{\dot A_\sigma=0} \,, \quad \sigma = x,y,z
\end{align}
or equivalently in the canonical coordinates the eigenvalues $-\Lambda^2$
of the three-dimensional real symmetric Hesse matrix 
$\partial^2 V_{\mathrm{h}}/\partial{\vec q}^2|_{\nabla V_{\mathrm{h}}=0}$
of the potential $V_{\mathrm{h}}({\vec q})$ are calculated.
The results are shown in Fig.~\ref{fig:stability}.
The eigenvalues appear in pairs of different
sign and are all pure imaginary for the stable state. The linear
stability for different trap frequencies shows qualitatively similar
behavior.

\subsection{Dynamics with a frozen Gaussian}
\label{sec:psos}
Hamilton's equations \eqref{eq:ham} describe the dynamics of the
soliton with the potential $V_\mathrm{h}(\vec q)$. The systematic
investigation and visualization of the dynamics of a Hamiltonian
system with three degrees of freedom is a nontrivial task. However,
the force in the $y$ direction is dominated by the strong harmonic
trap potential. If therefore $\gamma_y$ is sufficiently large at least
for the ground state of the condensate $q_y$ takes nearly the value of
the harmonic oscillator ground state. As a further
simplification of the problem we therefore restrict the dynamics to
the plane given by the condition
\begin{align}
  \ddot q_y = \dot p_y \approx \frac{1}{4q_y^3}-4\gamma_y^2 q_y
  \stackrel{!}{=} 0 \; \Rightarrow\; q_y = \frac{1}{2\sqrt{\gamma_y}} \, ,
\label{eq:frozen-gaussian}
\end{align}
which corresponds to a frozen Gaussian ansatz in the $q_y$ direction.
For $\gamma_y=2000$ the plane $q_y \approx 0.011$ is marked in
Fig.~\ref{fig:potential}(b).

\begin{figure}
\includegraphics[width=1\columnwidth]{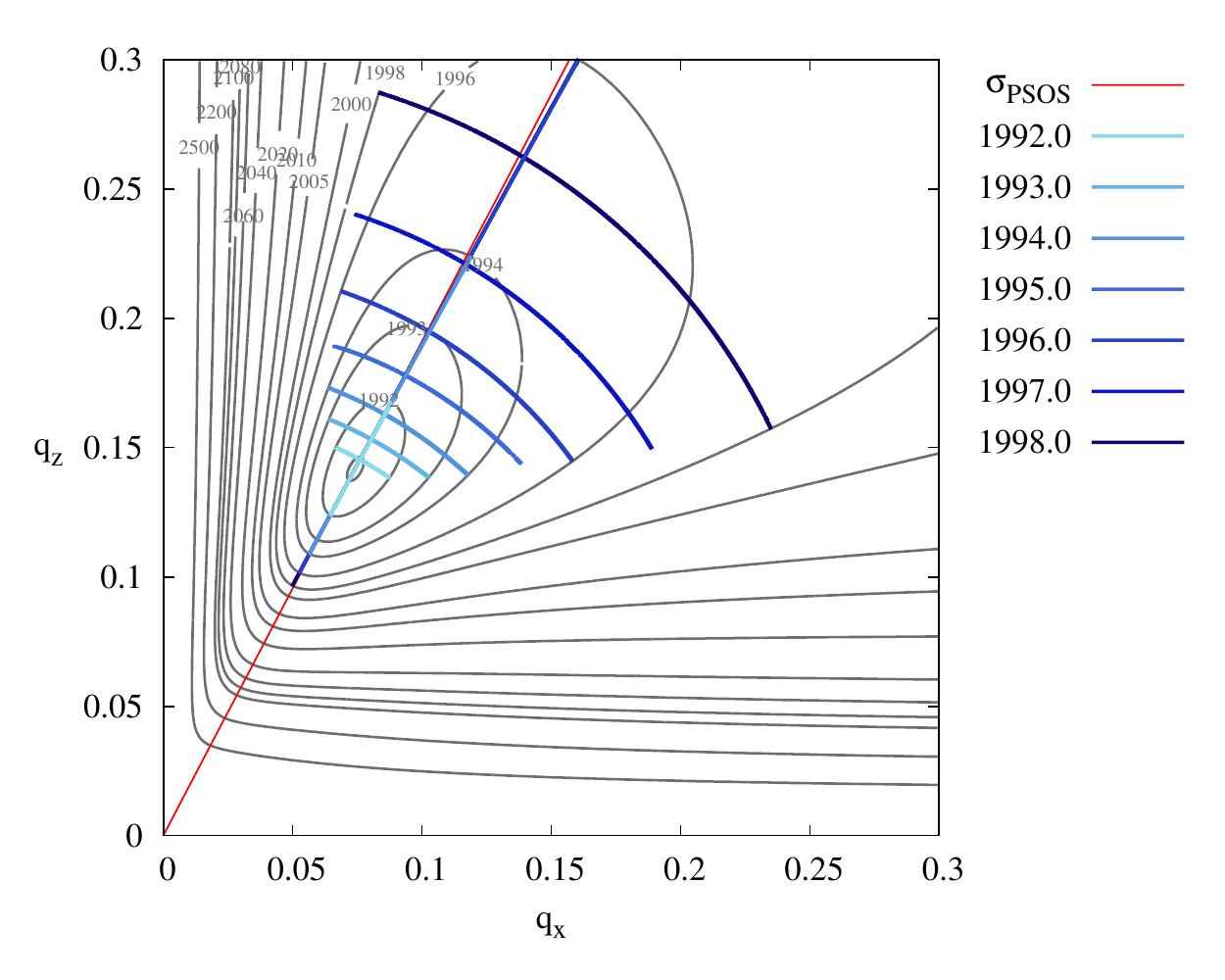}
  \caption{(Color online)
    Periodic trajectories in the potential
    $V_{\mathrm{2D,h}}$ at different values of energy for
    $\gamma_y=2000$ and scattering length $a=0.1$. The red line
    denotes the surface of section for the PSOSs. There are two
    periodic trajectories which belong to symmetric and antisymmetric
    oscillations of the soliton. The symmetric oscillations are hardly
    visible lying on top of one another and almost parallel to the
    surface of section.}
\label{fig:periodic1gauss}
\end{figure}

The dynamics of the wave function \eqref{eq:1gauss_ansatz} with fixed 
parameters $A_y^r=0$, $A_y^i=1/8q_y^2=\gamma_y/2$ is described in the
canonical coordinates $q_x$ and $q_z$ by the two-dimensional potential
\begin{align}
\label{eq:FrozenGaussian_V2D,h}
 &V_{\mathrm{2D,h}}\left( q_x,q_z \right) = \frac{1}{8q_x^2} + \frac{1}{8q_z^2}
  + \gamma_y + \frac{\sqrt{2\gamma_y}a}{4\sqrt{\pi}q_xq_z}  \\
 &+ \frac{1}{24\sqrt{2\pi}q_z}\left(-\frac{2\sqrt{\gamma_y}}{q_x}
  + \frac{1}{q_z^2} R_D\left( \frac{q_x^2}{q_z^2},\frac{1}{4q_z^2 \gamma_y},1
    \right)\right) \, , \nonumber
\end{align}
which is illustrated in Fig.~\ref{fig:periodic1gauss}.

To describe the dynamics of this two-dimensional system it is convenient 
to analyze Poincar\'e surfaces of section (PSOS).
An adequate choice of the PSOS shown in Fig.~\ref{fig:periodic1gauss}
is using the rotated coordinates and momenta
\begin{subequations}
\begin{align}
\begin{pmatrix}
q_1\\
q_2
\end{pmatrix} &=
\begin{pmatrix}[rr]
\cos \alpha & \sin \alpha \\
-\sin \alpha & \cos \alpha
\end{pmatrix}
\begin{pmatrix}
q_x\\
q_z
\end{pmatrix} \; ,\\
\begin{pmatrix}
p_1\\
p_2
\end{pmatrix} &=
\begin{pmatrix}[rr]
\cos \alpha & \sin \alpha \\
-\sin \alpha & \cos \alpha
\end{pmatrix}
\begin{pmatrix}
p_x\\
p_z
\end{pmatrix}
\end{align}
\label{eq:PSOS_coordinates}
\end{subequations}
with $\alpha = \arctan(q_{z,\min}/q_{x,\min})$ and the crossing condition 
$q_2 =0$. 
For constant energy different initial conditions are integrated.
The border of the allowed energy range is given by
\begin{align}
 p_1 = \pm \sqrt{2\left( E_{\mathrm{mf}}-V_{\mathrm{2D,h}} \right) } \, .
\end{align}
\begin{figure}
\includegraphics[width=0.95\columnwidth]{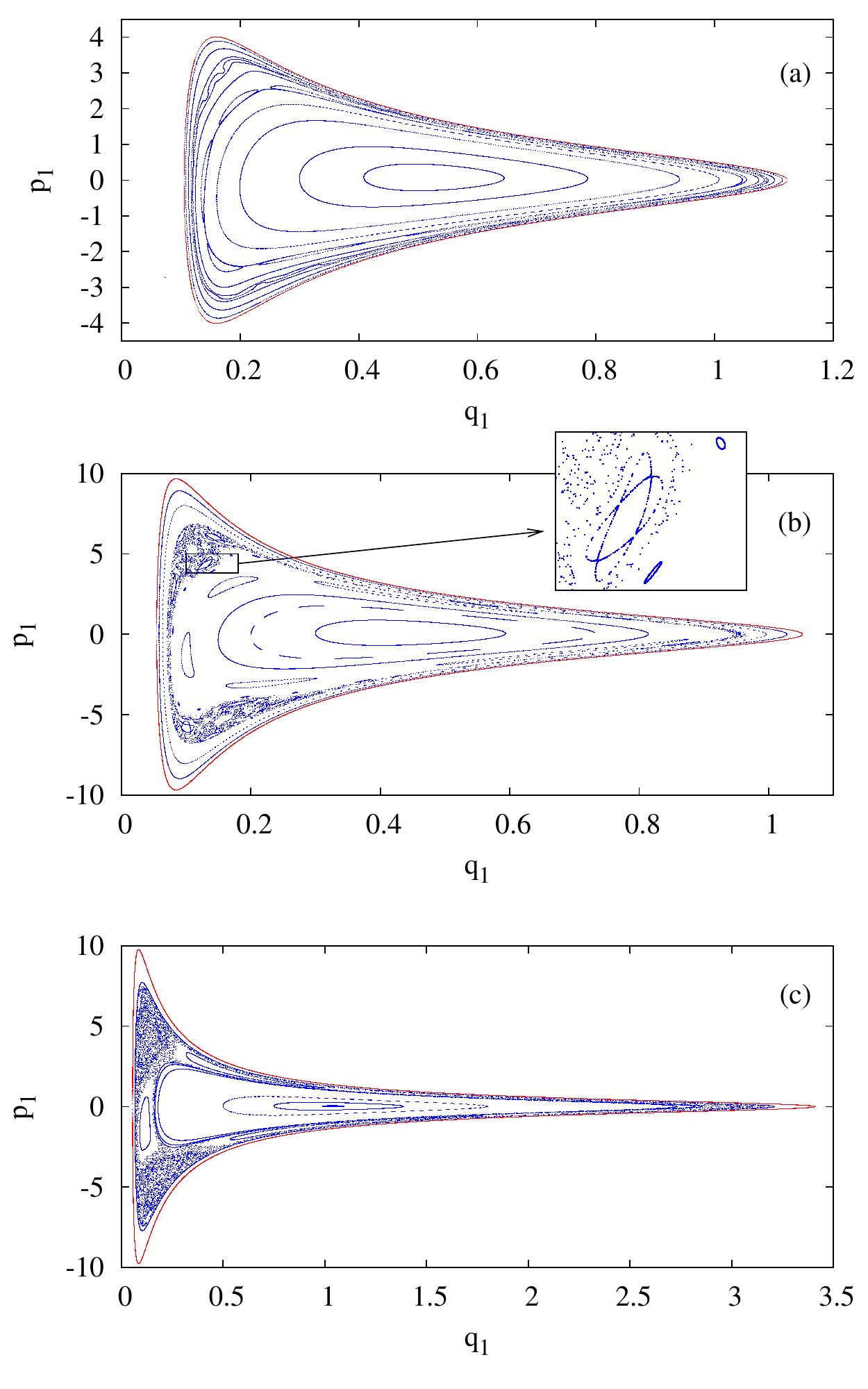}
\caption{(Color online)
 Poincar\'e surfaces of sections for the dynamics of the two-dimensional
 potential $V_{\mathrm{2D,h}}$ in Eq.~\eqref{eq:FrozenGaussian_V2D,h}
 in the rotated coordinates defined in Eq.~\eqref{eq:PSOS_coordinates}
 with trap frequency $\gamma_y=2000$ and
 (a) $a=0.1$, $E_{\mathrm{mf}}=1999.5$;
 (b) $a=0.08$, $E_{\mathrm{mf}}=1995$;
 (c) $a=0.08$, $E_{\mathrm{mf}}=1999$.
 In (a) the motion is completely regular.  Lowering the scatting length
 towards the bifurcation point chaotic dynamics appears as shown in (b).
 Increasing the mean-field energy close to $E_\mathrm{mf}=\gamma_y$ at
 constant scattering length enlarges the chaotic regions in the
 Poincar\'e map, see (c).}
\label{fig:PSOS}
\end{figure}
In Fig.~\ref{fig:PSOS}(a) a PSOS at an energy close to the ground
state is shown.
The motion is completely regular and an elliptic fixed
point of the Poincar\'e map is present, which belongs to the
antisymmetric periodic oscillation of the condensate. The symmetric
oscillation is not visible in the PSOS for the surface of section
being almost parallel to it. Increasing the energy towards
$E_\mathrm{mf}=\gamma_y$, the turning points of the periodic
oscillations in Fig.~\ref{fig:periodic1gauss} move to larger values of
$q_x$ and $q_z$ (for the symmetric oscillation only one turning point,
for the antisymmetric both). For $E_\mathrm{mf}=\gamma_y$ they lie at
infinity, i.e.\ the soliton dissolves. In the Poincar\'e map the
resonant tori decay according to the Poincar\'e-Birkhoff therorem
building the same number of elliptic and hyperbolic fixed points in
the Poincar\'e map. The closer to the bifurcation point the scattering
length is, the lower the energy is, where this takes place. This can
be seen especially in Fig.~\ref{fig:PSOS}. Further increasing of the
energy leads to regions of chaotic oscillations while the elliptic
fixed point moves outwards to larger values of $q_1$.

In the picture with a frozen Gaussian the solitons dissolve at energy
$E_{\mathrm{mf}}=\gamma_y$ in agreement with
\cite{Tikhonenkov08a,Santos09}. If this were always true it would
imply that in an experiment with realistic parameters (e.g., a
condensate with 10000 particles at $\gamma_y=2000$ and $a=0.1$) the
solitons must be cooled down to temperatures of about
$T=0.15\,\mathrm{\mu K}$ because the energy gap between the ground
state and the threshold $E_{\mathrm{mf}}=\gamma_y$ is very small (for
a more detailed discussion see Sec.~\ref{sec:res_coupled}). However, a
dynamical stabilization of the solitons at energies $E_{\mathrm{mf}} >
\gamma_y$ is possible as will be discussed for an ansatz with a single
Gaussian in the next Section \ref{sec:three_dynam} and for coupled
Gaussians in Sec.~\ref{sec:dyn_coupled}.

\subsection{Three-dimensional dynamics}
\label{sec:three_dynam}
In the frozen Gaussian approximation any excitation energy of the soliton
must be completely deposited in the $(q_x,q_z)$ motion, which leads to 
the dissolving of the soliton at the low threshold $E_{\mathrm{mf}}=\gamma_y$.
In this section we demonstrate that for the fully three-dimensional 
dynamics with the ansatz (\ref{eq:1gauss_ansatz}) a large amount of 
excitation energy can be stored in the $q_y$ motion dominated by the 
one-dimensional harmonic trap.

\begin{figure}
\includegraphics[width=0.95\columnwidth]{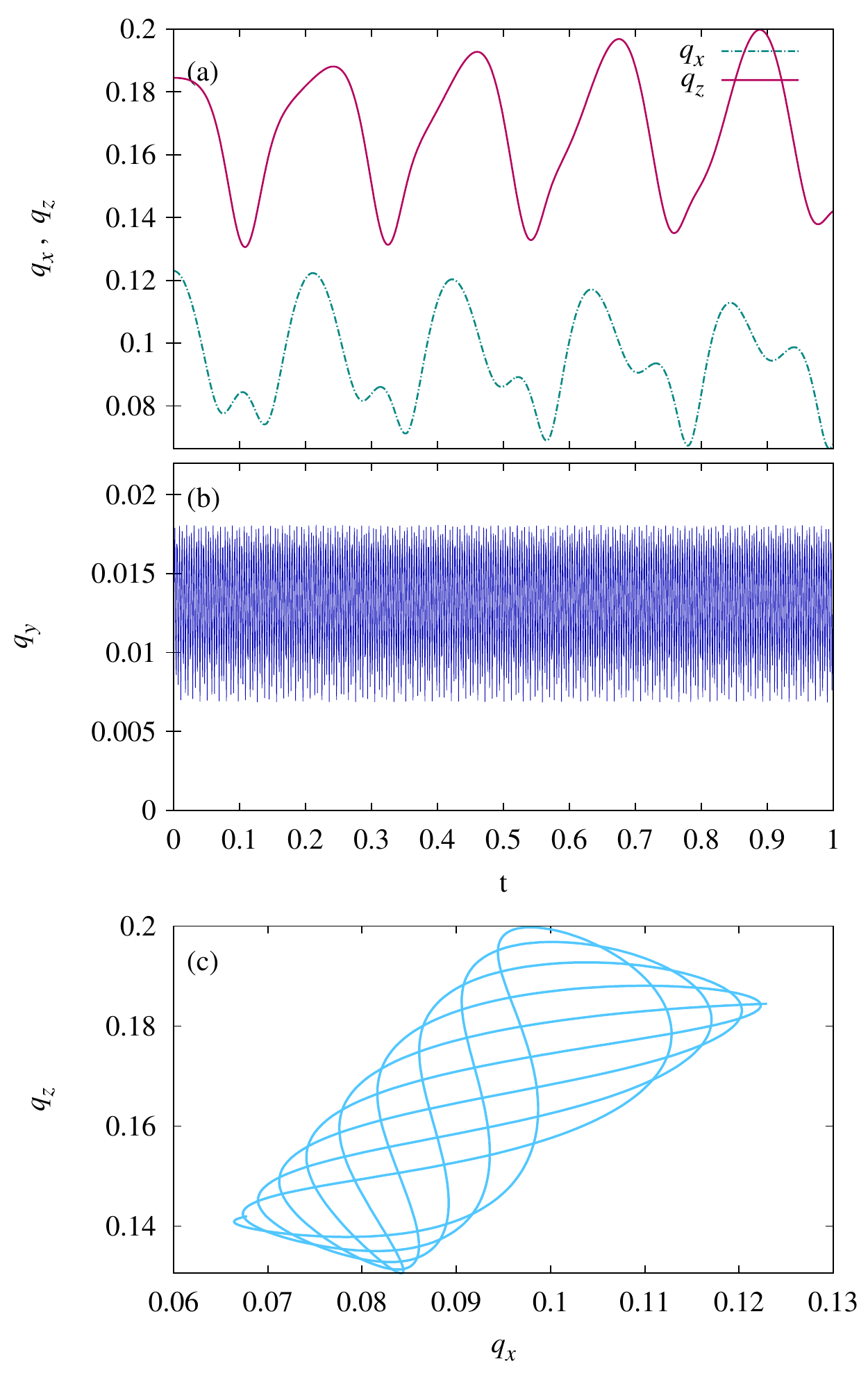}
\caption{(Color online)
 Three-dimensional trajectory of a soliton at trap frequency $\gamma_y=2000$,
 scattering length $a=0.1$, and at mean-field energy $E_\mathrm{mf}=3000$
 far above the threshold $E_{\mathrm{mf}}=\gamma_y$ where dissolving is
 possible.  (a) and (b): Time dependence of the expectation values 
 $q_\sigma(t)$ with $\sigma=x,y,z$.  (c) Lissajous type motion of
 the projection in the $(q_x,q_z)$ plane.}
\label{fig:dyn_1gauss}
\end{figure}

An example trajectory for trap frequency $\gamma_y=2000$, scattering length 
$a=0.1$, and with mean-field energy $E_\mathrm{mf}=3000$ far above the 
threshold at $\gamma_y$ is presented in Fig.~\ref{fig:dyn_1gauss}.
In (a) and (b) the expectation values
\begin{align}
 q_\sigma = \sqrt{\frac{\erw{\sigma^2}}{2}}
 = \frac{1}{8A^i_\sigma} \; , \quad \sigma = x,y,z
\label{eq:defqxqyqz}
\end{align}  
are drawn as functions of time. The slow oscillations in $q_x(t)$ and
$q_z(t)$ generate the Lissajous type motion of the quasi-2d soliton
visualized in Fig.~\ref{fig:dyn_1gauss}(c). The fast oscillation in
$q_y(t)$ results dominantly from the external harmonic trap. The
calculations with a single Gaussian thus indicate that a rather highly
excited soliton may be dynamically stabilized and does not dissolve in
contrast to the calculations with a frozen Gaussian
\cite{Tikhonenkov08a,Santos09}. However, it must be clarified (in
Sec.~\ref{sec:dyn_coupled}) whether the dynamical stabilization is
also possible with coupled Gaussians.

\section{Ansatz with coupled Gaussians}
\label{chap:coupledgauss}
Though the ansatz made in Sec.~\ref{chap:singlegauss} offers a
descriptive analysis of the soliton, comparison with calculations
where the GPE is solved numerically on a grid show that the results
for the ground state only hold qualitatively. As it will be shown, the
results for the dynamics of the soliton are qualitatively different as
well. To gain quantitatively correct results the ansatz for the trial
function has to be modified. As shown in \cite{Rau10a,Rau10b} the
condensate wavefunction can be well described
by a superposition of Gaussians.
We therefore apply the TDVP to coupled Gaussian wave packets
\cite{McLachlan1964a,Heller75a,Fab09a,Fab09b}.
The ground state will be determined by the imaginary time evolution (ITE)
with the norm conservation embedded in the TDVP as a constraint.

In this section we elaborate the theory and derive the formulae for the 
computations with coupled Gaussians.
The results for the stationary states and the dynamics of the solitons
are presented in Sec.~\ref{sec:results_coupled}.

\subsection{Time-dependent variational principle for GWPs}
\label{sec:TDVP}
In this article the TDVP provides the basis for the solution of the
time-dependent GPE. Its application to GWPs was originally introduced
by Heller \cite{heller:4979,Heller81a} for the description of atomic
and molecular quantum dynamics and later on applied to BEC
\cite{Rau10a,Rau10b}. For the convenience of the reader we here
shortly recapitulate this method.

The quantity
\begin{align}
  I = \vectornorm{\mathrm{i} \phi - H
    \chi(t)}^{2}\stackrel{!}{=}\min 
\label{eq:McLachlan_I}
\end{align}
is to be minimized with respect to $\phi$. Afterwards it is set $\phi
\equiv \dot \chi$. The wave function shall be parameterized by the
variational parameters $\vec z(t)$. The variation of $I$ in
Eq.~\eqref{eq:McLachlan_I} carries over to the variation of
$\vec z$ and $\dot{\vec z}$ and results in
\begin{align}
  \delta I &= \braket{\delta\dot{\chi}}{\dot{\chi}} +
  \braket{\dot{\chi}}{\delta\dot{\chi}} - \braket{\ii \delta
    \dot{\chi}}{H\chi} - \braket{H\chi}{\ii\delta\dot{\chi}}\nonumber \\
  &= \braket{\del{\chi}{\vec{z}} \delta\dot{\vec{z}}}{\dot{\chi}+\ii
    H\chi} + \braket{\dot{\chi}+\ii
    H\chi}{\del{\chi}{\vec{z}}\delta\dot{\vec{z}}}\,,
\label{eq:deltaI}
\end{align}
for the wave function $\chi$ being constant. Because the variational
parameters are complex quantities both brackets in
Eq.~\eqref{eq:deltaI} have to vanish separately. This yields
\begin{align}
\braket{\del{\chi}{\vec{z}}}{\del{\chi}{\vec{z}}\dot{\vec{z}}} &=
-\ii\braket{\del{\chi}{\vec{z}}}{H\chi}\,,
\end{align}
which can be written shortly as
\begin{align}
K\cdot \dot{\vec{z}} &= -\ii \vec{h}\,,\label{eq:McLachlan_2.32_Kdotz=-ih}
\end{align}
where $K$ is an hermitian positive definite matrix. The linear system
\eqref{eq:McLachlan_2.32_Kdotz=-ih} has to be solved for every time
step to integrate the equations of motion $\dot{\vec z}=f(\vec z)$.

We now choose a linear superposition of $N$ Gaussians as the trial function
\begin{align}
\chi = \Psi= \sum\limits_{k=1}^{N} g\left( \vec z^k,\vec{x} \right) = \sum\limits_{k=1}^{N}\ee{\ii\left(
\vec{x}A^k\vec{x} + \gamma^k \right) } \equiv \sum_{k=1}^{N} g^k
\label{eq:GWPansatz}
\end{align}
where $A^k = A^{k,r} + \ii A^{k,i}$ are complex diagonal matrices of
dimension $3\times 3$ and $\gamma^k=\gamma^{k,r}+\ii\gamma^{k,i}$
denotes the relative phases and the amplitude of the Gaussians,
respectively. The Gaussians are fixed at the origin. The time
evolution can be considered as the motion in an effective
time-dependent harmonic potential
\begin{align}
V_\mathrm{eff} \left( \vec x \right)= v_0 + \frac{1}{2} \vec{x} V_2 \vec{x}\,.
\end{align}
The dynamics of the GWPs are now determined by the TDVP, which
variationally fits the effective time-dependent harmonic potential
coefficients $v_0,V_2$ to the underlying potential. Splitting the
Hamiltonian $H=T+V$ and operating on the trial wave function
\eqref{eq:GWPansatz} yields
\begin{align}
  \ii \dot \Psi - T \Psi &=\sum\limits_{k=1}^{N}
    \Bigl[ \underbrace{- \dot \gamma^k +2\ii \Tr A^k}_{v_0^k}   \nonumber \\
  &+ \vec x \underbrace{\left( -\dot A^k - 4 \left( A^k \right)^2
    \right)}_{\frac{1}{2}V_2^k}\vec x \Bigr]\,g(\vec z^k,\vec x)\,.
\label{eq:TDVP+GWPeffPot}
\end{align}
The parameters $v_0^k$ are scalars and the matrices $V_2^k$ are
diagonal matrices. The equations of motion then read
\begin{subequations}\label{eq:TDVP+GWP_eom}
\begin{align}
\dot \gamma^k &= 2 \ii \Tr A^k - v^k_0\,, \\
\dot A^k &= -4\left( A^k \right)^2 - \frac{1}{2} V^k_2 \,.
\end{align}
\end{subequations}
The coefficients $v_0^k$ and $V_2^k$ have to be calculated from the
TDVP
\begin{align}
  \braket{\del{g(\vec z^k,\vec x)}{\vec{z}^k}}{\ii\dot{\Psi}- H \Psi } = 0\,,
\end{align}
\begin{align}
  \braket{x_\alpha^2 g(\vec z^l,\vec x)}{\sum\limits_{k=1}^{N} \left(
  v^k_0 + \frac{1}{2} \vec x V^k_2 \vec x \right)\,g(\vec z^k,\vec x)} = 0 \,.
\label{eq:systembraket}
\end{align}
Combining the coefficients $v_0^k$ and $V_2^k$ in a complex vector
$\vec v$ the set of equations \eqref{eq:systembraket} can be written
as
\begin{align}
  K\vec v = \vec r\,.
\label{eq:McLachlan_II}
\end{align}
The positive definite hermitian matrix $K$ includes in contrast to
Eq.~\eqref{eq:McLachlan_2.32_Kdotz=-ih} the kinetic operator.
Eq.~\eqref{eq:McLachlan_II} has to be solved for every time step and
the resulting vector $\vec v$ has then to be inserted in
Eq.~\eqref{eq:TDVP+GWP_eom}.

In the numerical integration of the equations of motion
\eqref{eq:TDVP+GWP_eom} (e.g. with a standard algorithm like
Runge-Kutta), the quadratic term can lead to the numerical overflow
\cite{heller:4979}. It is therefore reasonable to split up the
matrices $A^k$ into two matrices $C$ and $B$ according to
$A^k=B^k(C^k)^{-1}$. For the case of diagonal matrices $A^k$, $C^k$
and $B^k$ are diagonal, too. Hence the splitting can be done
component-by-component. This leads to
\begin{align}
\dot C^k =4B^k\; ; \quad \dot B^k = -\frac{1}{2} C^k V_2^k\,,
\end{align}
where $C(0)=\mathbbm{1}$ and $B(0) = A(0)$ are taken as initial values.

\subsection{Constraints in the TDVP}
\label{sec:constraints}
The most important reason for the introduction of constraints is the
conservation of the norm in the imaginary time evolution
(cf.\ Sec.~\ref{sec:ite}). Apart from that inequality constraints can
be used to avoid matrix singularities which arise from overcrowding
the basis set \cite{Fabcic08a}. In principle $m$ constraints can be
introduced combined in the $m$-dimensional vector $\vec f$. The
constraints are embedded then by a set of Lagrangian multiplicators
$\vec \lambda \in \mathbbm R^m$
\begin{align}
L=I+\vec \lambda \bar M \dot{\bar{\vec z}}\,,
\end{align}
where $\bar M= \del{\vec f}{\bar{\vec z}}$ is a real $m\times 2n_p$
matrix. To find the minimum of $I$
\begin{align}
\del{L}{\vec \omega}=0\quad \text{with} \quad \vec \omega \equiv
\begin{pmatrix}
\dot{\vec z}^r\\
\dot{\vec z}^i\\
\vec \lambda
\end{pmatrix} =
\begin{pmatrix}
\dot{\bar{\vec z}}
\end{pmatrix} \in \mathbbm{R}^{2n_p+m}
\end{align}
has to be calculated which yields the system of equations
\begin{align}
\begin{pmatrix}
\begin{array}{c|c}
\bar{K}^{\strut} & \bar{M}^T \\
\hline
\bar{M}^{\strut} & 0
\end{array}
\end{pmatrix}
\begin{pmatrix}
\dot{\bar{\vec z} }\\ \vec{\lambda}
\end{pmatrix}			&=
\begin{pmatrix}
\vec{\bar{h} }\\ 0
\end{pmatrix},
\label{eq:constraint1}
\end{align}
with
\begin{align}
\bar{K} &=
\begin{pmatrix}[rr]
K^r & -K^i \\
K^i &  K^r
\end{pmatrix},\;
\vec{\bar{h} } =
\begin{pmatrix}[rr]
\vec{h}^i \\
-\vec{h}^r
\end{pmatrix}\,.\nonumber
\end{align}
In the special case of Gaussian wave packets the equations of motion
(split into real and imaginary part and combining the variational
parameters to 
$\bar{\vec z} =(\gamma^{1,r} \dots \gamma^{N,r}, A^{1,r} \dots 
A^{N,r}, \gamma^{1,i} \dots \gamma^{N,i}, A^{1,i} \dots A^{N,i})$
can be written in the form
\begin{align}
\dot{\bar{\vec z}} = \tilde{U} \bar{\vec v} + \tilde{\vec d}\,,
\end{align}
where $\tilde U$ consists of the terms linear in $\vec v$ and the
constant terms are absorbed in the vector $\tilde{\vec d}$. The
constraints combined in $\vec f =(f_1,\dots ,f_n)$ yield
\begin{align}
  \dot{\vec f}=\del{\vec f}{\bar{\vec z}}\tilde U \bar{\vec v} +
  \del{\vec f}{\bar{\vec z}}\tilde{\vec d} \equiv \bar U \bar{\vec v}
  + \bar{\vec d} = 0\,.\label{eq:constraint-dotf=Uv+d}
\end{align}
The set of linear equations \eqref{eq:constraint1} then reads
\begin{align}
\begin{pmatrix}
\begin{array}{c|c}
\bar{K}^{\strut} & \bar{U}^T \\
\hline
\bar{U}^{\strut} & 0
\end{array}
\end{pmatrix}
\begin{pmatrix}
\bar{\vec v}\\ \vec{\lambda}
\end{pmatrix} &=
\begin{pmatrix}[rr]
\bar{\vec r}\\ -\bar{\vec d}
\end{pmatrix},
\end{align}
with
\begin{align}
\bar{K} &=
\begin{pmatrix}[rr]
K^r & -K^i \\
K^i &  K^r
\end{pmatrix},\;
\bar{\vec{r}} =
\begin{pmatrix}[rr]
\vec{r}^i \\
\vec{r}^r
\end{pmatrix}\,.
\end{align}
Solving this system of linear equations and inserting the result
vector in Eq.~\eqref{eq:TDVP+GWP_eom} enables one to integrate the
equations of motion.

\subsection{Energy functional}
\label{sec:Emf}
To gain the mean-field energy of the soliton the time-dependent
variational parameters of the GWPs are used in the evaluation of the
integrals in the energy functional. The energy functional for $N$
coupled Gaussians reads
\begin{align}
E_{\mathrm{mf}} &= \matrixel{\Psi}{-\Delta}{\Psi } +
\matrixel{\Psi}{V_\mathrm{t}}{\Psi }\nonumber \\ &+
\frac{1}{2}\left(\vphantom{\frac{1}{2}}
\matrixel{\Psi}{V_\mathrm{c}}{\Psi }+
\matrixel{\Psi}{V_\mathrm{d}}{\Psi }\vphantom{\frac{1}{2}}
\right)\,.
\label{eq:Emf_coupled}
\end{align}
With $\sigma=x,y,z$ the integrals in (\ref{eq:Emf_coupled}) yield
\begin{align}
&\matrixel{\Psi}{-\Delta}{\Psi} =
 \sum\limits_{l,k} \matrixel{g^l}{-\Delta}{g^k} \\ &\quad= \sum\limits_{l,k}\left( \left[ -2\ii
\sum\limits_\sigma \left( A_\sigma^k + \frac{\left( A^k_\sigma \right)^2}{{A_\sigma^{l}}^* - A^k_\sigma}
\right) \right] \braket{g^l}{g^k} \right)\,,\nonumber
\end{align}
\begin{align}
\matrixel{\Psi}{V_\mathrm{t}}{\Psi} &= -\frac{1}{2} \ii
\sum\limits_{l,k,\sigma}\frac{\gamma_\sigma^2}{{A_\sigma^{l}}^*-A_\sigma^k } \braket{g^l}{g^k}\,,
\end{align}
\begin{align}
\matrixel{\Psi}{V_\mathrm{c}}{\Psi} 
= 8\pi a \sum\limits_{l,k,i,j} & \Bigg[ \left(\prod\limits_\sigma
\sqrt{\frac{\ii \pi}{A_\sigma^k+A_\sigma^i-{A_\sigma^{l}}^*-{A_\sigma^{j}}^* } }
\,\right)\nonumber \\
&\times \ee{\ii \left( \gamma^k+\gamma^i-{\gamma^{l}}^*-{\gamma^{j}}^*\right) }
\Bigg]\,,
\end{align}
\begin{align}
&\matrixel{\Psi}{V_\mathrm{d}}{\Psi} = 
\frac{4\pi}{3} \sum\limits_{i,j,k,l}  \Bigg[ \left(\prod\limits_{\sigma} \sqrt{ \frac{\ii
\pi}{A_\sigma^k+A_\sigma^i-{A_\sigma^l}^*-{A_\sigma^j}^* }}\, \right)\nonumber\\
&\times \left[
\tilde\kappa_x\tilde\kappa_y\,R_D\left( \tilde\kappa_x^2,\tilde\kappa_y^2,1 \right) -1\right] \,
 \ee{\ii \left( \gamma^k+\gamma^i-{\gamma^{l}}^*-{\gamma^{j}}^*\right)}\Bigg]\, ,
\end{align}
with the abbreviations
{\small
\begin{align*}
\tilde\kappa_x &= \sqrt{\frac{ \left( A_x^i+A_x^k- {A_x^j}^* - {A_x^l}^* \right) \left( A_z^i - {A_z^j}^*
\right)
\left( A_z^k -{A_z^l}^* \right) }%
{ \left( A_x^i -{A_x^j}^* \right)\left( A_x^k - {A_x^l}^* \right)
\left( A_z^i+A_z^k -{A_z^j}^* -{A_z^l}^* \right) }  }
\end{align*} }
and
{\small
\begin{align*}
\tilde\kappa_y &= \sqrt{\frac{ \left( A_y^i+A_y^k- {A_y^j}^* - {A_y^l}^* \right) \left( A_z^i - {A_z^j}^*
\right)
\left( A_z^k -{A_z^l}^* \right) }%
{ \left( A_y^i -{A_y^j}^* \right)\left( A_y^k - {A_y^l}^* \right)
\left( A_z^i+A_z^k -{A_z^j}^* -{A_z^l}^* \right) }  }\,.
\end{align*} }
The additional integrals in the TDVP $\expect{\sigma^2V(\vec x)}$,
$\expect{\sigma^4V(\vec x)}$ and $\expect{\alpha^2\beta^2V(\vec x)}$
are obtained easily from the integrals in the energy functional.

\subsection{Imaginary time evolution}
\label{sec:ite}
In principle the ground state of the system of coupled Gaussians could
be obtained by a nonlinear root search in the equations of motion.
Since $4N$ complex initial values are required, it is difficult to
find the fixed points this way for a large number of Gaussians $N$.
The imaginary time evolution represents an alternative for the
determination of the ground state.

\begin{figure}
\includegraphics[width=1\columnwidth]{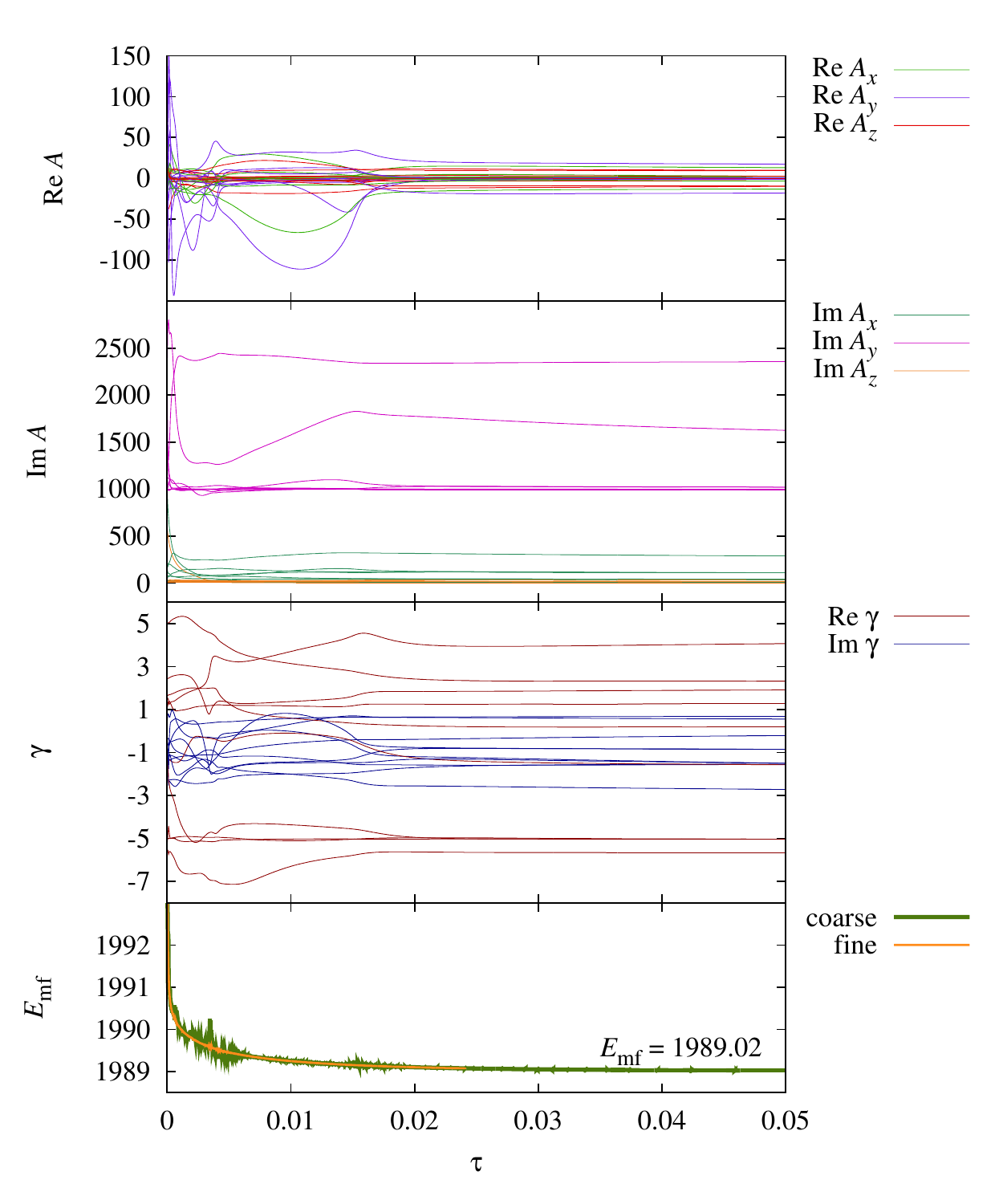}
  \caption{(Color online)
    Imaginary time evolution of 10 coupled Gaussians. The
    parameters used are $\gamma_y = 2000$, $a=0.1$. Plotted are, from
    the upper to the lower panel, the real part of the width
    parameters, the imaginary part of the width parameters, real and
    imaginary part of $\gamma$ and the mean-field energy. The
    mean-field energy converges to the ground state of the system as
    described in the text. For comparison the mean-field energy in a
    fast calculation with coarse integration accuracy is shown as
    well.}
\label{fig:ite}
\end{figure}
In the linear Schr\"odinger equation the transformation $t \rightarrow
\tau = \ii t$ leads to an exponential damping of the excited states
and converges to the ground state for $\tau \rightarrow \infty$. The
method can be applied to the nonlinear GPE as well. In imaginary time
the norm is not conserved any more. This is crucial in the nonlinear
GPE because the decay of the norm has the meaning of particle losses.
To avoid the norm decay we implement the norm conservation as a
constraint
\begin{align}
f &= \sum_{k,l} \underbrace{\braket{g^l}{g^k}}_{f_{kl}} \stackrel{!}{=} 1
\label{eq:constraint}
\end{align}
in the TDVP as elaborated above. Supposing a general complex rotation
of the time $\tau=\ee{\ii \Omega}t$ with $\xi = \xi^r+\ii \xi^i =
-\ee{-\ii \Omega}$ the equations of motion can be written in the
general form
\begin{align}
\dot{\bar{\vec z}} =\tilde{U} \bar{\vec v} + \vec{\tilde{d}} 
\end{align}
with
\begin{align*}
\tilde{U} &=
\begin{pmatrix}
\left( \xi^r \right)_{N\times N}	&	\left( 0\right)_{3N\times N}	&	\left(
-\xi^i\right)_{N\times N}	&	\left( 0\right)_{3N\times N}	\\
\left( 0\right)_{N\times 3N}	&	\left( \xi^r\right)_{3N\times 3N}	&	\left(
0\right)_{N\times 3N}	&	\left( -\xi^i\right)_{3N\times 3N}	\\
\left( \xi^i\right)_{N\times N}	&	\left( 0\right)_{3N\times N}	&	\left(
\xi^r\right)_{N\times N}	&	\left( 0\right)_{3N\times N}	\\
\left( 0\right)_{N\times 3N}	&	\left( \xi^i\right)_{3N\times 3N}	&	\left(
0\right)_{N\times 3N}	&	\left( \xi^r\right)_{3N\times 3N}	\\
\end{pmatrix}\,,
\end{align*}
\begin{align*}
\vec{\tilde{d}} &=
\xi^r
\begin{pmatrix}
2\,\Tr{A^{k,i}} \\ 4\left( A^{k,r} \right)^2 -4 \left( A^{k,i} \right)^2 \\ -2\,\Tr{A^{k,r}} \\ 8A^{k,r}
A^{k,i}
\end{pmatrix}
  \nonumber\\
 &+ \xi^i
\begin{pmatrix}
2\,\Tr{A^{k,r}} \\ -8A^{k,r} A^{k,i} \\ 2\,\Tr{A^{k,i}} \\ 	4\left( A^{k,r} \right)^2 -4 \left( A^{k,i}
\right)^2
\end{pmatrix}
\,, \quad
\bar{\vec v} =
\begin{pmatrix}
\vec v^{r}_{0}	\\	\frac{1}{2} \vec V^{r}_{2}	\\	\vec v^{i}_{0}	\\	\frac{1}{2}
\vec V^{i}_{2}
\end{pmatrix}\,,
\end{align*}
where the abreviative notation in $\tilde U$ can be understood as
follows: Every entry $(\lambda )_{a\times b}$ is an $a\times b$
submatrix $\tilde U'$. If $\lambda =0$ all elements of $\tilde U' =0$,
otherwise $\tilde U'=\lambda\cdot \mathbbm 1$. The vectors $\vec
v^r_0=(v^{1,r}_0,\dots,v^{N,r}_0)^T$, $\vec
v^i_0=(v^{1,i}_0,\dots,v^{N,i}_0)^T$, $\vec
V^r_2=(V_2^{1,r},\dots,V_2^{N,r})^T$ and $\vec
V^i_2=(V_2^{1,i},\dots,V_2^{N,i})^T$ contain the real and imaginary
parts of the coefficients of the effective potential. The entries in
$\tilde{\vec d}$ are vectors with the index $k$ running from $1$ to
$k$. One special case is the real time evolution $\xi^r=1$, $\xi^i=0$
and the other special case the imaginary time evolution $\xi^r=0$,
$\xi^i=1$. The gradient vector $\bar{F}^T=\partial f / \partial \bar
z$ in Eq.~\eqref{eq:constraint-dotf=Uv+d} is given by
\begin{subequations}
\begin{align}
\del{f}{\gamma^{k,r}} &=  -2 \sum\limits_{l=1}^N \Imag f_{kl}\,,\\
\del{f}{A^{k,r}_\sigma} &= -\sum\limits_{l=1}^N \Imag \frac{f_{kl} }{a^{kl}_\sigma }\,,\\
\del{f}{\gamma^{k,i}} &= -2 \sum\limits_{l=1}^N \Real f_{kl}\,,\\
\del{f}{A^{k,i}_\sigma} &= -\sum\limits_{l=1}^N \Real \frac{f_{kl}}{a^{kl}_\sigma }\,,
\end{align}
\end{subequations}
with which one obtains $\bar{U} = \bar{F}^T\cdot \tilde{U}$ and
$\bar{\vec d}=\bar{F}^T\cdot \tilde{\vec d}$.

With this method the norm is conserved during the numerical
integration for rather long times. However for very long times a small
drift in the norm is present due to the numerical error of the
integration. The ITE of a system of 10 coupled Gaussians is shown in
Fig.~\ref{fig:ite}.

If the relative accuracy of the integrater is chosen coarse, the
mean-field energy does not decay monotonously but for long times it
converges to the same value as an integration with a fine relative
accuracy thus a coarse accuracy can be chosen for the sake of time.
The ITE is very robust to the initial choice of the wave function.
This also holds for large numbers of Gaussians coupled, where the
nonlinear root search often fails. However, only the stable ground
state is accessible with this method.

\section{Results with coupled Gaussians}
\label{sec:results_coupled}
Applying the variational ansatz with coupled Gaussians to the wave
functions of dipolar condensates we are now able to obtain
quantitatively correct ranges of the parameters where stable quasi-2d
solitons can exist. For an experimental realization of solitons with
chromium atoms realistic trap frequencies are typically a few hundred
Hertz for condensates with about $10000$ to $20000$ particles. These
values roughly correspond to an interval of $2000 \lesssim \gamma_y
\lesssim 20000$. 
We therefore focus our calculations on two values for
the scaled trap frequency, viz.\ $\gamma_y=2000$ and $\gamma_y=20000$.

\subsection{Stationary ground state}
\label{sec:res_coupled}
\begin{figure}
\includegraphics[width=0.95\columnwidth]{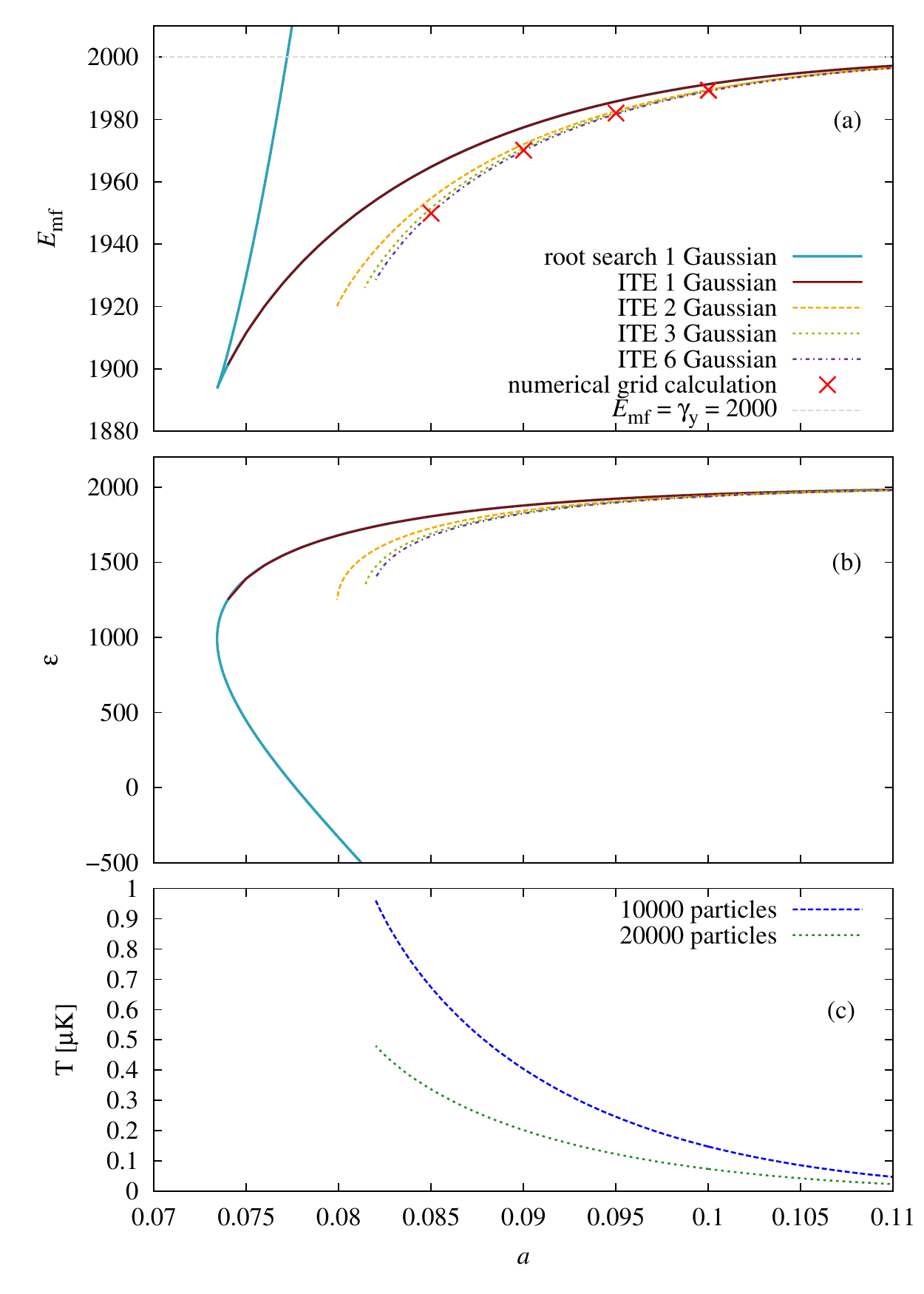}
\caption{(Color online) (a) Mean-field energy as a function of the
  scattering length of the ground state of a soliton at trap
  frequency $\gamma_y =2000$ obtained by the ITE with different
  numbers of Gaussians. For comparison the result for a single
  Gaussian, obtained by a nonlinear root search and the results
  obtained by numerical grid calculations are plotted as well. (b) The
  chemical potential for the same parameters as above. (c) Temperature
  for different particle numbers belonging to the difference of upper
  limit $E_\mathrm{mf}=\gamma_y$ and ground state energy.}
\label{fig:coupled1}
\end{figure}

\begin{figure}
\includegraphics[width=0.95\columnwidth]{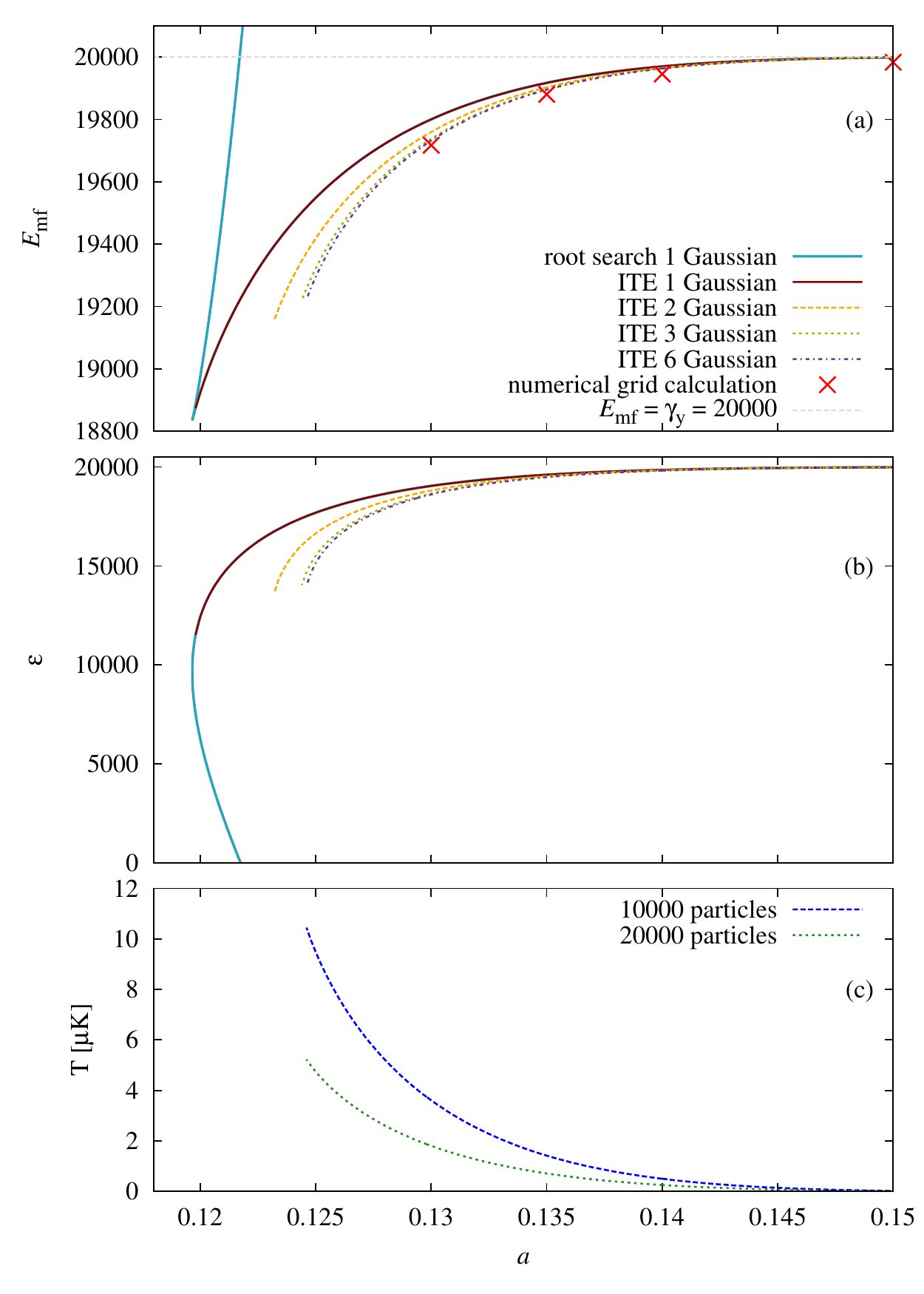}
\caption{(Color online)
 Same as Fig.~\ref{fig:coupled1} but for trap frequency $\gamma_y = 20000$.
 Compared to the results for $\gamma_y=2000$ the range of the 
 scattering length is smaller but the temperature is higher.}
\label{fig:coupled2}
\end{figure}

\begin{figure}
\includegraphics[width=\columnwidth]{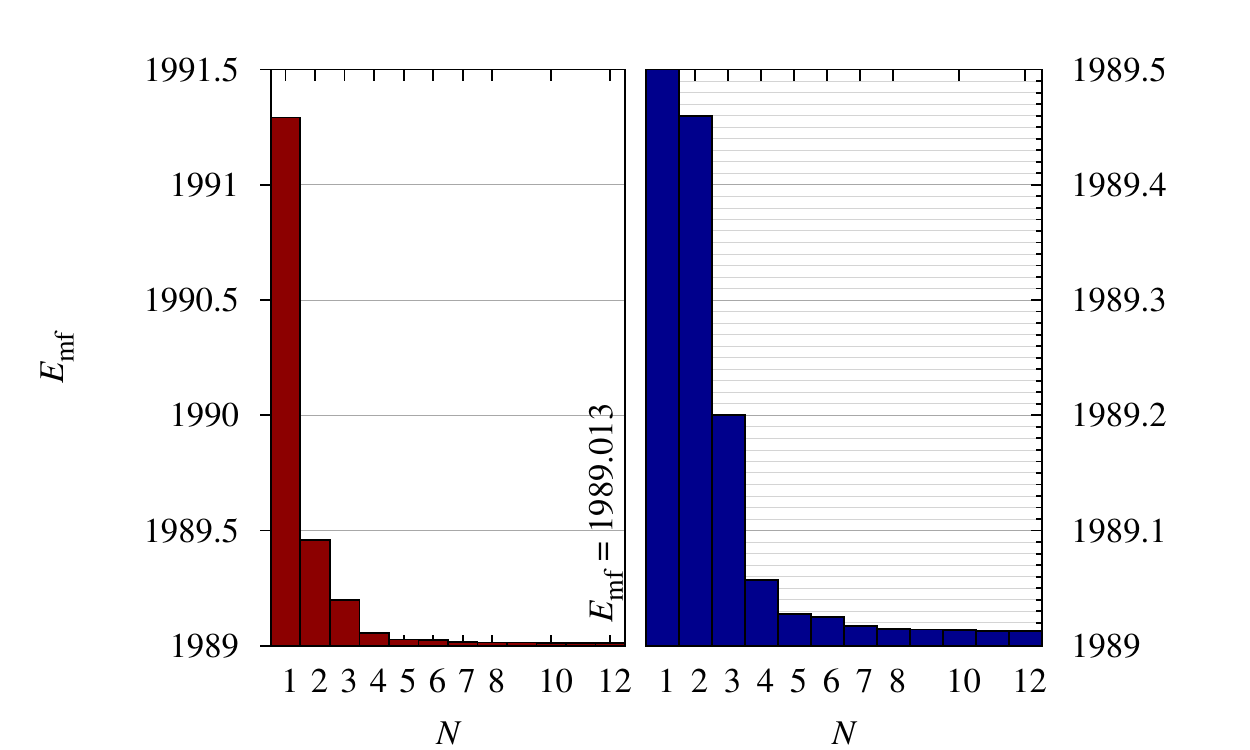}
\caption{(Color online)
 Mean-field energy for trap frequency $\gamma_y=2000$ and fixed scattering
 length $a=0.1$.  The mean-field energy converges fast with increasing
 number of Gaussians.}
\label{fig:coupled3}
\end{figure}

In Figs.~\ref{fig:coupled1} and \ref{fig:coupled2} the results 
for the ground state of the soliton are shown at trap frequencies
$\gamma_y=2000$ and $\gamma_y=20000$, respectively.
In Figs.~\ref{fig:coupled1}(a) and \ref{fig:coupled2}(a) the
mean-field energy as a function of the scattering length is
plotted for up to 6 coupled Gaussians. For comparison the calculation
using a nonlinear root search for one Gaussian as discussed in
Sec.~\ref{sec:stat_1g} is displayed, as well. The mean-field energy
and also the chemical potential [see Figs.~\ref{fig:coupled1}(b) and
\ref{fig:coupled2}(b)] converge with increasing number of Gaussians
$N$. The detailed convergence properties of the calculations with
increasing number of Gaussians are illustrated in
Fig.~\ref{fig:coupled3}, where calculations for up to 12 Gaussians are
shown at constant scattering length $a=0.1$.
The ground state energy converges fast with the number of Gaussians
and the corrections above a number of about 5 Gaussians can be
neglected. The small energy differences between the variational and
the grid calculations in Fig.~\ref{fig:coupled2}(a) may be due to the
finite grid size, which is limited by an acceptable computation time.

In the calculations with a single Gaussian a stable and an unstable
state are created in a tangent bifurcation at a scattering length
$a=a_{\mathrm{cr}}$. Using coupled Gaussians the critical scattering
length where the condensate collapses is shifted to higher values,
e.g., from $a_{\mathrm{cr}}^{N=1}=0.0734$ to
$a_{\mathrm{cr}}^{N=6}=0.0820$ at trap frequency $\gamma_y=2000$ and
from $a_{\mathrm{cr}}^{N=1}=0.1197$ to $a_{\mathrm{cr}}^{N=6}=0.1246$
at trap frequency $\gamma_y=20000$. The ITE can only provide the
stable ground state but no unstable states which are certainly also
involved in the bifurcation. For dipolar condensates in an
axisymmetric trap a complicated bifurcation scenario has been revealed
\cite{Rau10,Rau10b}, and it will be an interesting future task to
study bifurcations of the soliton states in more detail.

The corrections to the calculations with a single Gaussian decrease
with growing scattering lengths. This can be understood when looking
at the spatial size of the soliton [cf.\ Fig.~\ref{fig:1gauss}(c)].
For large expansion of the condensate the nonlinear interaction terms
in the GPE are small and thus the mean-field energy is dominated by
the ground state energy of the harmonic trap in the $y$ direction.
Above a certain threshold value for the scattering length the soliton
is no longer bound but dissolves. The thresholds at, e.g., $a=0.137$
for $\gamma_y=2000$ and $a=0.157$ for $\gamma_y=20000$ obtained with a
single Gaussion (see Fig.~\ref{fig:1gauss}) are only very slightly
shifted to higher values when using coupled Gaussians.

The extended variational approach allows us to compute for each trap
frequency an accurate lower and upper critical value of the scattering
length, i.e., a range where stable solitons in dipolar BECs can exist.
However, in an experiment thermal excitations of the soliton may cause
the dissolving of the soliton, and therefore it is also necessary to
determine an energy or temperature limit for the existence of stable
solitons. As a dissolving condensate without any interaction must have
at least the zero point energy $E_\mathrm{mf}=\gamma_y$ of the
harmonic trap in the $y$ direction the difference between this
threshold and the mean-field energy of the ground state,
\begin{align}
 \Delta E = \gamma_y-E_{\mathrm{mf}}^{\mathrm{g}} \approx \frac{3N k_B T}{2E_\dd}
\label{eq:DE}
\end{align}
can be taken for a very conservative estimation of the temperature 
at which the soliton should survive thermal excitations.
In Figs.~\ref{fig:coupled1}(c) and \ref{fig:coupled2}(c) that temperature 
is plotted as a function of the scattering length for two condensates 
with $N=10000$ and $N=20000$ particles.
With this conservative estimate a soliton at, e.g., trap frequency 
$\gamma_y=2000$ and scattering length $a=0.1$ must be cooled down to
about $T=0.15\,\mu K$.
However, solitons may exist at much higher temperatures when they are
dynamically stabilized as discussed in Sec.~\ref{sec:three_dynam} for
the ansatz with a single Gaussian.
We now extend that discussion to the ansatz with coupled Gaussians.

\subsection{Dynamics with coupled Gaussians}
\label{sec:dyn_coupled}
The ansatz with a single Gaussian used to describe the dynamics of 
the solitons in Sec.~\ref{sec:three_dynam} and especially the further
simplification with a frozen Gaussian in Sec.~\ref{sec:psos} are 
basically mathematical model systems, and the results obtained
with these models have to be seen from an academic point of view.
For a realistic description it is necessary to investigate the 
dynamics of the solitons with the extended ansatz \eqref{eq:GWPansatz} 
of coupled Gaussians which has shown to significantly improve the 
results for the ground state in Sec.~\ref{sec:res_coupled}.
Of special interest is to search for the existence of dynamically 
stabilized solitons at energies above the threshold $E_{\mathrm{mf}}=\gamma_y$.
To this aim the equations of motion \eqref{eq:TDVP+GWP_eom} are 
integrated with the initial wave function chosen appropriately to 
yield the desired mean-field energy.

The condensate wave function \eqref{eq:GWPansatz} with $N$ coupled 
Gaussians is parametrized by $8N$ real time-dependent variational 
parameters.
For the visualization of the wave function we use the reduced set 
of parameters
\begin{align}
 q_\sigma = \sqrt{\frac{1}{2}\langle\Psi|\sigma^2|\Psi\rangle} \; ,
 \quad \sigma = x,y,z
\end{align}
which are related to the extension of the soliton in the three spatial 
dimensions and can be directly compared with the results for a single 
Gaussian using the coordinates in Eq.~\eqref{eq:defqxqyqz}.

\begin{figure}
\includegraphics[width=0.95\columnwidth]{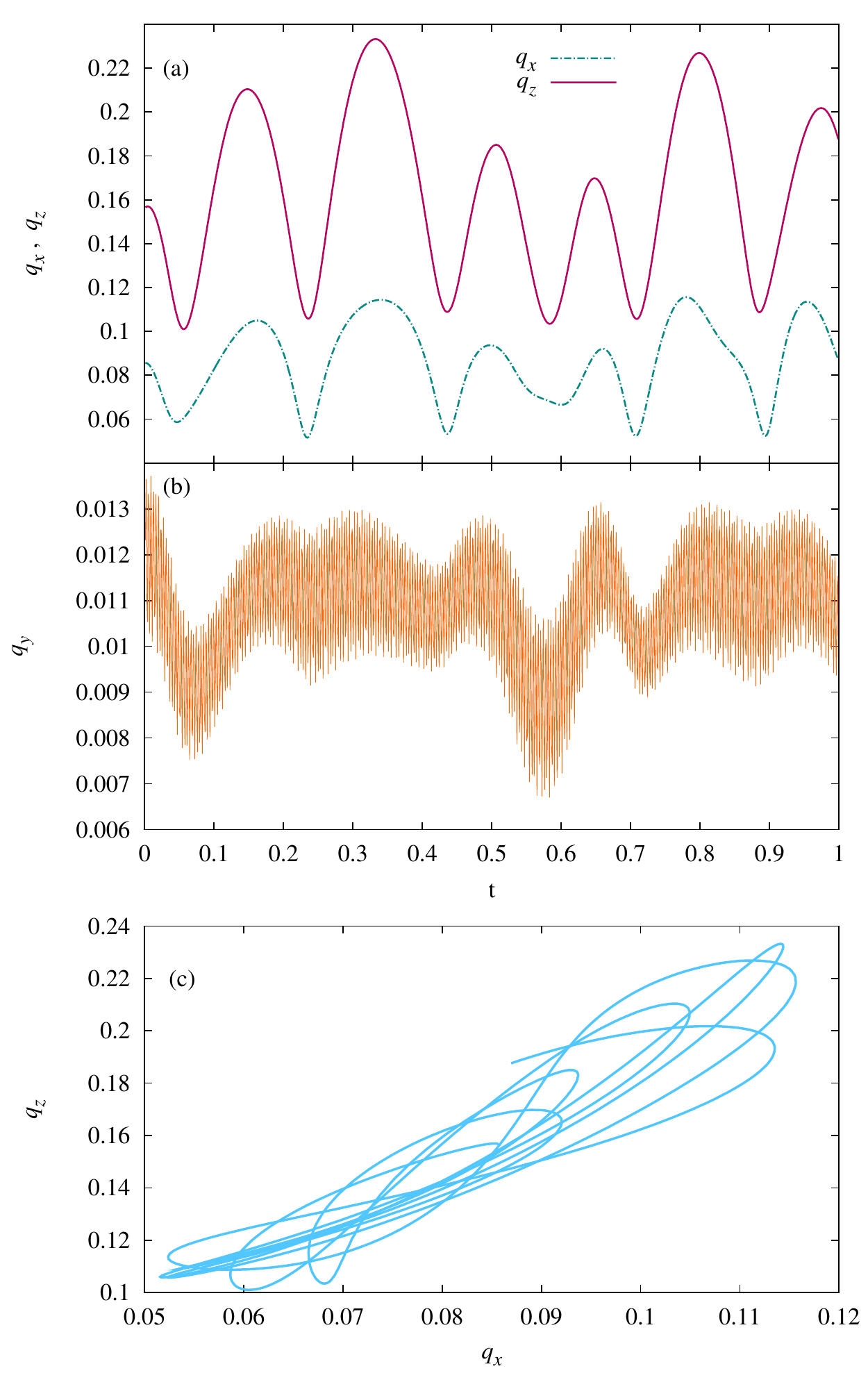}
\caption{(Color online) Dynamics with 2 coupled Gaussians for trap
  frequency $\gamma_y=2000$, scattering length $a=0.1$, and
  mean-field energy $E_{\mathrm{mf}}=2100$. In (a) the parameters
  $q_x$ and $q_z$ and in (b) the parameter $q_y$ in the trap direction
  are shown as functions of time. (c) Projection of the trajectory in
  the $(q_x,q_z)$ plane. The soliton does not dissolve although the
  mean-field energy is quite far above the threshold at
  $E_{\mathrm{mf}}=\gamma_y$.}
\label{fig:dyn_2gauss}
\end{figure}

For energies close to the ground state energy periodic oscillations
can be found. The parameters of the turning points of the periodic
oscillations show a rather complicated behavior when the mean-field
energy is increased. The different periodic oscillations vanish in
multiple bifurcations close to the ground state energy. Above
$E_\mathrm{mf}=\gamma_y$ no periodic oscillations were found. However
increasing the energy above $E_\mathrm{mf}=\gamma_y$ yields
oscillations of the soliton which do not destroy it. Applying a frozen
Gaussian approximation yields stable oscillations only slightly above
this limit. But allowing oscillations in all spatial directions
stabilizes the condensate and gives rise to an oscillating soliton as
shown in Fig.~\ref{fig:dyn_2gauss} for a trajectory at trap frequency
$\gamma_y=2000$, scattering length $a=0.1$, and mean-field energy
$E_{\mathrm{mf}}=2100$.

The excitation energy of the dynamically stabilized non-dissolving
soliton in Fig.~\ref{fig:dyn_2gauss} is $\Delta
E=E_{\mathrm{mf}}-E_{\mathrm{mf}}^{\mathrm{g}}=111.0$ which for a
condensate with $10000$ particles corresponds to a temperature of
about $T=1.5\,\mathrm{\mu K}$. For the system with $\gamma_y=2000$ and
$a=0.1$ stable oscillations can be found up to $E_\mathrm{mf}=2400$.
This is about $40$ times the energy difference in Eq.~\eqref{eq:DE}
and yields for an estimation of the temperature
$\mathrm{T}=5.5\,\mathrm{\mu K}$ for $N=10\,000$ particles.
Calculations with a higher number of coupled Gaussians and with
various initial conditions show that the oscillating soliton is
stable. The choice of the initial wave function is not crucial for the
stability of the condensate.

\section{Conclusion and outlook}
\label{sec:conclusion}
We investigated anisotropic quasi-2d solitons in dipolar Bose-Einstein
condensates with the time-dependent variational principle using both
the descriptive ansatz of a single Gaussian wave function and coupled
Gaussians to calculate the ground state of the system. The lower limit
of the energy obtained with a superposition of Gaussians is in full
agreement with numerical grid calculations. For a given trap frequency
$\gamma_y$ we are able to give boundaries for the scattering length
where stable solitons can exist. The energy gap between the mean-field
energy of the ground state and the threshold energy where the soliton
can dissolve is typically quite small. However, the investigation of
the dynamics of the dipolar BECs have revealed the existence of
dynamically stabilized non-dissolving solitons at energies far above
the threshold $E_{\mathrm{mf}}=\gamma_y$, and opens the possibility to
create solitons at experimentally accessible temperatures
\cite{Griesmaier05a}. This discovery may thus stimulate experiments on
solitons in dipolar Bose-Einstein condensates.

\begin{acknowledgments}
  R.E.\ is grateful for support from the Landesgraduiertenf\"orderung of
  the Land Baden-W\"urttemberg.
\end{acknowledgments}

%

\end{document}